\documentclass[a4paper,fleqn,usenatbib,useAMS]{mn2e}

\usepackage{graphicx}	
\usepackage{aas_macros}
\usepackage{amsmath}	
\usepackage{amssymb}	
\usepackage{multicol}   
\usepackage{multirow}
\usepackage{bm}		    
\usepackage{pdflscape}	
\usepackage{gensymb}
\usepackage{epstopdf}
\usepackage{hyperref}
\usepackage{xcolor}
\usepackage{afterpage}
\usepackage{float}

\usepackage[T1]{fontenc}
\usepackage{ae,aecompl}

\usepackage{txfonts}

\usepackage{makecell}

\title[]{A candidate field for deep imaging of the Epoch of Reionization observed with MWA}

\author[]{Xueying Zhang$^{1}$, Qian Zheng$^{1}$\thanks{E-mail: qzheng@shao.ac.cn}, Linhui Wu$^{1}$\thanks{wulinhui@shao.ac.cn}, Quan Guo$^{1}$\thanks{guoquan@shao.ac.cn}, Stefan~W.~Duchesne$^{2}$, \newauthor Mengfan He$^{1, 3}$, Huanyuan Shan$^{1}$, Xiang-ping Wu$^{4}$, Melanie Johnston-Hollitt$^{5}$, \newauthor Feiyu Zhao$^{1, 6}$, Qingyuan Ma$^{7}$\\
$^{1}$Shanghai Astronomical Observatory, Chinese Academy of Sciences 80 Nandan Road, Shanghai 200030, China \\
$^{2}$CSIRO Space and Astronomy, PO Box 1130, Bentley WA 6102, Australia \\
$^{3}$Department of Astronomy, Tsinghua University, Beĳing 100084, China \\
$^{4}$National Observatories of China, 20A Datun Road, Beijing, China \\
$^{5}$Curtin Institute for Data Science, GPO Box U1987, Perth, WA 6845, Australia \\
$^{6}$University of Chinese Academy of Sciences, No.1 yanqihu East Road, Beijing 101408, China \\
$^{7}$ School of Physics and Astronomy, Sun Yat-sen University, Zhuhai 519082, China
}
\date{Last updated 2018 January 31; in original form 2017 XXX X}

\pubyear{2025}

\begin{document}
\label{firstpage}
\pagerange{\pageref{firstpage}--\pageref{lastpage}}
\maketitle

\begin{abstract}
Deep imaging of structures from the Cosmic Dawn (CD) and the Epoch of Reionization (EoR) in five targeted fields is one of the highest priority scientific objectives for the Square Kilometre Array (SKA). Selecting ‘quiet’ fields, which allow deep imaging, is critical for future SKA CD/EoR observations. Pre-observations using existing radio facilities will help estimate the computational capabilities required for optimal data quality and refine data reduction techniques.
In this study, we utilize data from the Murchison Widefield Array (MWA) Phase II extended array for a selected field to study the properties of foregrounds. We conduct deep imaging across two frequency bands: 72–103 MHz and 200–231 MHz. We identify up to 2,576 radio sources within a 5-degree radius of the image center (at RA (J2000) 8$^h$, Dec (J2000) 5$^\circ$), achieving approximately 80\% completeness at 7.7 mJy and 90\% at 10.4 mJy for 216 MHz, with a total integration time of 4.43 hours and an average RMS of 1.80 mJy. Additionally, we apply a foreground removal algorithm using Principal Component Analysis (PCA) and calculate the angular power spectra of the residual images. Our results indicate that nearly all resolved radio sources can be successfully removed using PCA, leading to a reduction in foreground power. However, the angular power spectra of the residual map remains over an order of magnitude higher than the theoretically predicted CD/EoR 21 cm signal. Further improvements in data reduction and foreground subtraction techniques will be necessary to enhance these results.

\end{abstract}





\section{Introduction}
\label{sec:introduction}

The redshifted 21 cm signal from neutral hydrogen is a crucial tool for probing the Dark Ages (redshift $z \approx 30 - 200$), Cosmic Dawn (CD, $z \approx 15 - 30$) and Epoch of Reionization (EoR, $z \approx 6 - 15$). There are three primary approaches to measure the CD/EoR by detecting this signal: assessing spatial fluctuations through statistical properties, conducting tomographic imaging of ionized structures, and measuring the sky-averaged global signal. The redshifted 21 cm signal can be detected in the low-frequency radio band from approximately 50 MHz to 200 MHz.
Many radio interferometers are under development or in operation, with the exploration of CD/EoR as a primary focus. Notable examples include the Murchison Widefield Array (MWA; \citealt{bowman2013science}; \citealt{tingay2013murchison}), the 21 Centimeter Array (21CMA; \citealt{zheng2016radio}), the Low-Frequency Array (LOFAR; \citealt{van2013lofar}), the Giant Metrewave Radio Telescope (GMRT; \citealt{paciga2013simulation}), the Precision Array for Probing the Epoch of Re-ionization (PAPER; \citealt{parsons2010precision}) and the Hydrogen Epoch of Reionization Array (HERA; \citealt{deboer2017hydrogen}) all dedicated to statistical measurements of the 21 cm signal. 
Additionally, experiments such as the Experiment to Detect the Global EoR Signature (EDGES; \citealt{bowman2008toward}; \citealt{bowman2018absorption}), the Sonda Cosmológica de las Islas para la Detección de Hidrógeno Neutro (SCI-HI; \citealt{voytek2014probing}), the Large Aperture Experiment to Detect the Dark Ages (LEDA; \citealt{bernardi2015foreground}), and the Shaped Antenna Measurement of the Background Radio Spectrum (SARAS; \citealt{patra2013saras}) aim to detect the global signal of CD/EoR. 
Detecting the CD/EoR is a key scientific project of the Square Kilometre Array (SKA), which has the capability to both statistically measure the 21 cm signal and directly image CD/EoR structures.
However, the redshifted 21 cm signal is obscured by foregrounds that are 4 to 5 orders of magnitude higher, presenting significant challenges for CD/EoR observations. In addition to the foregrounds, systematic errors, such as incomplete sampling, inaccurate calibration, and confusion limits, pose major difficulties in using the SKA for tomographic imaging of ionized structures.

For MWA EoR observations, there are three selected EoR fields called EoR0, EoR1 and EoR2, each covering a sky region of about $20^{\circ}\times 20^{\circ}$. 
Many works about data reduction techniques and statistic measurements are published with the data observed on these three EoR fields and upper limits of the CD/EoR detection are given (\citet{procopio2017high}; \citet{barry2019improving}). \citet{trott2020deep} measured the limits on the power spectrum of deep observations over the three MWA EoR fields. \citet{rahimi2021epoch} presented the EoR power spectrum limits from MWA targeted at EoR1 field, and compared the results with EoR0 field. 
For LOFAR observations, two fields are selected for CD/EoR study, one is the North Celestial Pole (NCP) field, and the other one is the 3C 196 field. A lot of works about LOFAR CD/EoR detections have been published (\citet{chapman2012foreground}; \citet{yatawatta2013initial}), upper limits of CD/EoR detection are presented for the LOFAR EoR fields. 
NCP region is also the sky field observed with 21CMA. The key science goal of 21CMA is CD/EoR detection. The ground-fixed antennas observe the NCP region 24 hours a day. 

In addition to the aforementioned sky fields, several other fields have also been selected as candidates for EoR detection. For instance, in \citet{zheng2020pre}, seven ideal fields in the southern sky are selected for deep imaging of the EoR. Large-scale diffuse structures are avoided in the selected fields. The selected candidate fields also exhibit the lowest average surface brightness and the lowest variance. However, the selection presented in \citep{zheng2020pre} is constrained by the source catalogues used; as the observation of radio sources at low frequencies becomes more complete, the results will also be improved. 
To successfully extract the faint HI 21cm signal from CD/EoR and image the EoR ionized structures with SKA, it is imperative to choose ‘quiet’ fields and comprehend the characteristics of the source distribution. This selection not only facilitates in predicting the potential of future deep CD/EoR observations but also provides a foundation for refining data reduction techniques, including imaging and sky model construction, statistical algorithms and foreground removal. Pre-observations of candidate fields with existing radio facilities, particularly MWA telescopes, are indispensable. These observations aid in estimating the computational resources required to attain the desired data quality and evaluating the computational efficiency for detecting the EoR.

To detect the faint EoR signals, we require a good understanding of the foregrounds, which necessitates enhancing our observational sensitivity to higher levels and improving the completeness of weak sources detection. This is not only essential for the calibration of low-frequency wide-field observations but also crucial for better subtracting foreground sources and obtaining the EoR signals.
Over the past few years, there are a number of low frequency radio observations and many radio source catalogues have been published. Source catalogues (\citet{zheng2016radio}; \citet{zhao2022north}) are published for understanding the properties of radio sources at low frequencies and improving data calibrations and foregrounds removals in the future 21CMA data reduction. 
\citet{franzen2016154} published the radio source counts at 154\,MHz with MWA observations, which is targeted on the MWA EoR0 field centered at [(RA, Dec) = (00:00:00, -27:00:00)]. Source catalogues from the GaLactic and Extragalactic All-sky MWA survey \citep[GLEAM;][]{wayth2015gleam} have been published \citep{hurley2017galactic}. GLEAM covers 72 - 231\,MHz and Declinations south of $+30^\circ$. In \citep{hurley2022galactic}, a source catalogue of 78,967 components are detected in 1,447 square degrees sky region covered by GLEAM-X DR1. 
Due to the improved observational sensitivity, the completeness of sources has been further enhanced. Therefore, we need to conduct deep imaging of candidate fields by improved telescopes to further enhance the completeness of weak sources detection.

Observing the selected fields with SKA1-Low precursors is helpful in building and improving data reduction pipelines and understanding the foregrounds and backgrounds properties. Therefore, we observed one selected sky field with MWA. MWA is a low-frequency radio interferometer located at the Murchison Radio-astronomy Observatory in Western Australia. This site is designated as a protected radio-quiet zone, where the SKA1-Low is also being constructed. \citep{Off15} provides a detailed discussion of local frequency interference. The operational frequency range of the MWA spans from 72 MHz to 300 MHz, with an observation frequency bandwidth of 30.72 MHz. The maximum baseline length of extended configuration is 5.3 km, while the angular resolution is about 1.3 arcmin at 154MHz \citep{wayth2018phase, beardsley2019science}. For more details about the telescope and its scientific applications, refer to \citet{wayth2018phase} and \citep{beardsley2019science}\footnote{\href{https://research.csiro.au/mro/inyarrimanha-ilgari-bundara/}{https://research.csiro.au/mro/inyarrimanha-ilgari-bundara/}}.

The center of observed field is very close to one of the selected fields presented in \citep{zheng2020pre}, so it can be considered as a ‘quiet’ field for future CD/EoR observations. In this work, we present the deep imaging results of the selected sky field and provide predictions for future SKA CD/EoR observations. With the observation conducted in an extended configuration, this study focuses on analysing the properties of foregrounds rather than extracting large-scale background signals. Data reduction pipelines introduced in \citep{Du20} are adopted in this work. Principal Component Analysis (PCA) is adopted for removing the foregrounds from our generated 7.68 MHz sub-band images, and angular power spectra of the residuals are calculated. After foreground subtraction, the angular power spectra of the residuals remain over an order of magnitude higher than the theoretically estimated CD/EoR background signal. Improvements in data reduction and foreground subtraction methods are necessary for achieving better results. 

The structure of this paper is outlined as follows: Section~\ref{sec:observation} introduces the observation of one selected sky field with MWA and data reduction. In Section~\ref{sec:sources}, we introduce the properties of the radio sources detected in the selected field. In Section~\ref{sec:catalogue}, we describe the catalogue properties. Section~\ref{sec:foregrounds} delves into the statistic measurements of the selected field. Finally, Section~\ref{sec:discussion} presents conclusions and discussion. Throughout this paper, we use the spatially flat $\Lambda$CDM model in our work with fiducial parameters $h = 0.67, \Omega_{\Lambda} = 0.6911, \Omega_m = 0.3089, \Omega_k = 0.0$ in agreement with the results from \citet{aghanim2020planck}.

\section{Observation and Reduction}
\label{sec:observation}

In this section, we describe the details of our one selected field observed by MWA, and the data reduction pipeline before creating the deep wide-band images.

\subsection{One Selected Field and MWA Observation}

In this work, we adopt the data from MWA project G0044 (PI: Qian Zheng). G0044 is proposed for deep imaging of two selected ‘quiet’ fields, [(RA, Dec) = (7:55:38.4, +5:51:36.0)] and [(RA, Dec) = (10:32:33.7, -12:39:36.0)], with the aim of preparing for future SKA CD/EoR observations. The telescope was in MWA Phase II extended configuration during the observation. One of the G0044 fields [(RA, Dec) = (7:55:38.4, +5:51:36.0)] covers one of the selected candidate fields presented in \citep{zheng2020pre}, so we adopt this field for analysis in this work and will refer to it briefly as the 'G0044 field' in the following content. This field was observed from April 8 to April 30, 2018. The observations were conducted across five frequency bands of the GLEAM survey, enabling effective avoidance of strong radio frequency interference (RFI) \citep{offringa2010post, offringa2012morphological}. In this study, we processed data from two specific frequency bands spanning 72–103 MHz and 200–231 MHz, centered at 87.675 MHz and 215.675 MHz, which we will refer to as the 88 MHz and 216 MHz bands, respectively. Each observation snapshot has a frequency bandwidth of 30.72 MHz, a duration of 120 seconds, a correlated frequency resolution of 10 kHz, and a correlated integration time of 0.5 seconds. We use a max number of 169 observation snapshots in the 88 MHz band and 133 observation snapshots in the 216 MHz band. 

\begin{figure*}
\begin{center}
 \includegraphics[width=17.0cm]{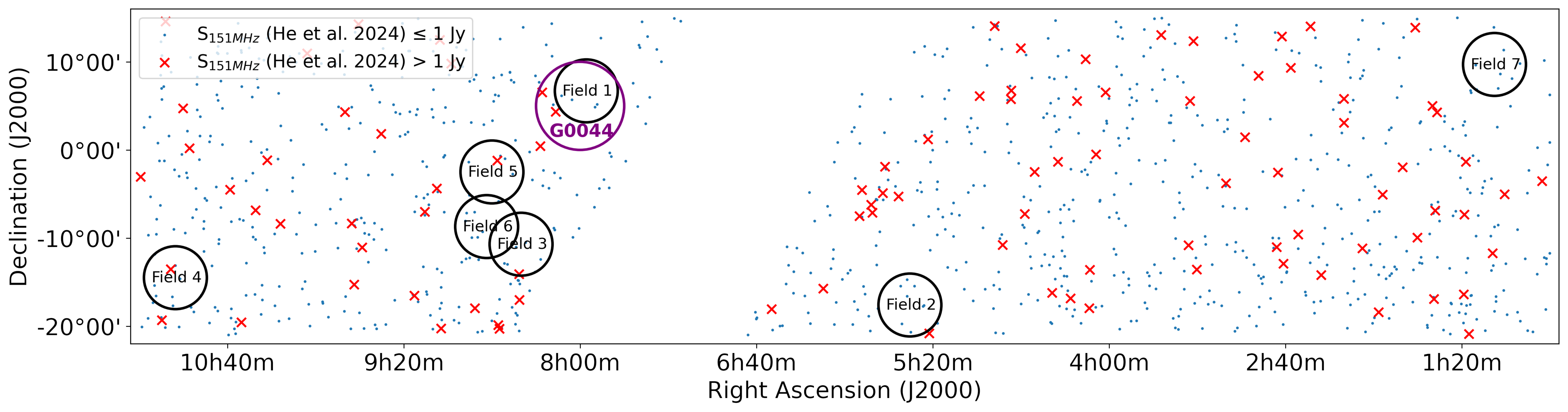}
 \caption{Distribution of spectral peak (SP) radio sources \citep{he2024influence}, especially showing bright SP sources with flux densities above 1 Jy. G0044 field (a purple circle covering a radius of 5 degrees) and 7 selected fields (black circles each covering 40 square degrees) are also shown with right ascension and declination listed in Table \ref{tab:7fields}. The G0044 field covers most region of Field 1.}
\label{fig:7fields}
\end{center}
\end{figure*}

The imaging center of G0044 field is located in one of the seven selected fields in \citet{zheng2020pre}, and the central coordinates of these fields are shown in Table \ref{tab:7fields}. We also plot these fields in Figure \ref{fig:7fields}, while the area of each selected fields is 40 degrees$^2$, or 3.57$^\circ$ in radius, and the region of G0044 field utilized for calculations in this work is 5$^\circ$ in radius. 
In Figure \ref{fig:7fields}, we also plot the distribution of spectral peak (SP) radio sources in \citet{he2024influence} based on the data from GLEAM survey, especially showing the bright SP sources with flux densities above 1 Jy marked with red crosses. Although the influence of SP sources can be ignored, the residuals from bright SP sources after foreground subtraction may still affect the final results of future CD/EoR deep imaging \citep{he2024influence}. Therefore, if possible, avoiding bright SP sources in the selected sky region can reduce the challenges in subsequent CD/EoR data reduction and signal extraction.

\begin{table}
	\centering
	\caption{The central right ascension (RA) and declination (Dec) of G0044 field and 7 selected candidate fields from Table 2 in \citet{zheng2020pre}.}
	\label{tab:7fields}
	\begin{tabular}{ccc}
	\hline
    Field  & RA        & Dec         \\
           & (J2000)   & (J2000)     \\
    \hline
    G0044  & 08:00:00  & +05:00:00   \\
    1      & 07:57:12  & +06:42:00   \\
    2      & 05:30:24  & -17:36:00   \\
    3      & 08:26:48  & -10:42:00   \\
    4      & 11:03:36  & -14:30:00   \\
    5      & 08:40:00  & -02:30:00   \\
    6      & 08:42:24  & -08:42:00   \\
    7      & 01:05:12  & +09:42:00   \\        
    \hline
	\end{tabular}
\end{table}

\subsection{Data Reduction}
\begin{figure*} 
\begin{center}
 \includegraphics[width=17.0cm]{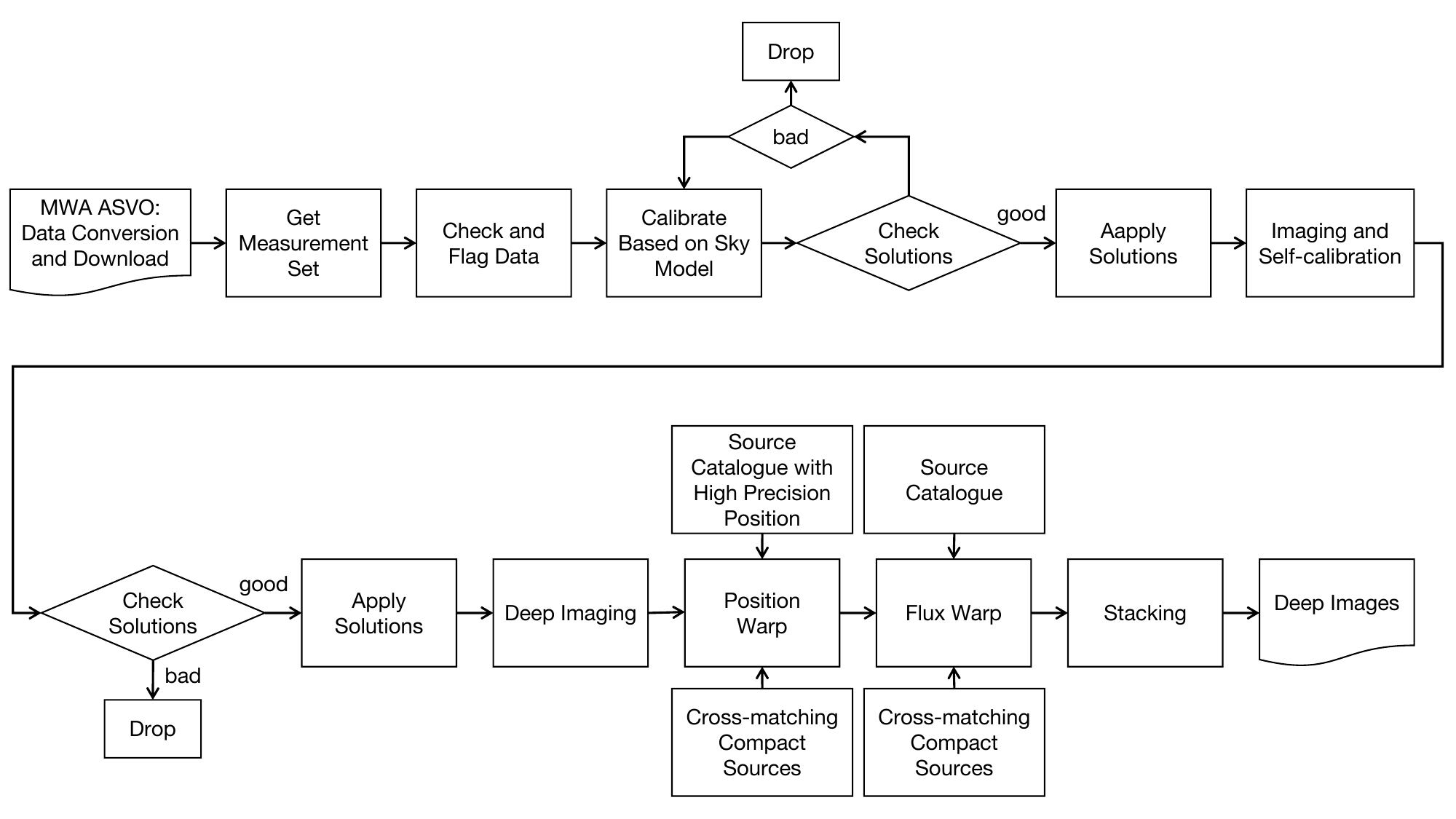}
 \caption{Illustration of the data reduction pipeline used in this work. Showing all the different steps from data conversion to calibration and deep imaging.}
\label{fig:pipeline}
\end{center}
\end{figure*}

The MWA observations for Project G0044 comprise hundreds of 2-minute snapshots. Data reduction mainly follows the procedures demonstrated in \citet{hurley2017galactic} and \citet{Du20}. We use the MWA All-Sky Virtual Observatory (ASVO)\footnote{\href{https://asvo.mwatelescope.org}{https://asvo.mwatelescope.org}} system to convert raw telescope products to the standard 'MeasurementSet' format using \texttt{COTTER} \citep{Off15} and perform preliminary radio frequency interference (RFI) flagging using \texttt{AOFlagger} \citep{offringa2012morphological}. The raw visibilities are averaged to have a frequency resolution of 40\,kHz, and a time resolution of 4\,s. These averaged visibilities are then calibrated based on the \texttt{Mitchcal} algorithm \citep{Off16} using a global sky model as described in \citet{Du20}. The imaging of each snapshot is conducted using \texttt{WSClean} \citep{Off14,Off17} with multiscale CLEANing. Self-calibration for each snapshot is also performed to further correct the residual amplitude and phase errors. Final imaging (Stokes I) is performed for each 2-min snapshot with "channels-out=4" and Briggs weighting ($\rm robust= 0.5$, see \citealt{Bri95}), obtaining both the 30.72 MHz bandwidth images and 4 subband images of 7.68 MHz bandwidth. Then, these images are corrected for astrometry to consider the position-dependent offsets are largely introduced by the ionosphere using \texttt{fits\_warp.py} \citep{hancock2018source}, and the position-dependent flux scale is set using \texttt{flux\_warp.py}\footnote{\href{https://gitlab.com/Sunmish/flux\_warp}{https://gitlab.com/Sunmish/flux\_warp}} \citep{Du20}.  Both of these tools take an input sky model generated by GLEAM, the 1.4 GHz NRAO\footnote{National Radio Astronomy Observatory} VLA\footnote{Very Large Array} Sky Survey \citep[NVSS;][]{condon1998nrao} and/or the Sydney University Molonglo Sky Survey \citep[SUMSS;][]{Bock1999,Mauch2003,Murphy2007}. This sky model is in turn cross-matched to point sources in the snapshot image catalogues to calculate astrometric off sets and flux density discrepancies. For detailed methods, please refer to \cite{Duchesne2021}. Finally, for each band and corresponding 4 subbands, their snapshot images are stacked to create mosaics as described in \citet{Du20}. The details of the pipeline are illustrated in Figure \ref{fig:pipeline}. 

\begin{figure*} 
\begin{center}
 \includegraphics[width=17.0cm]{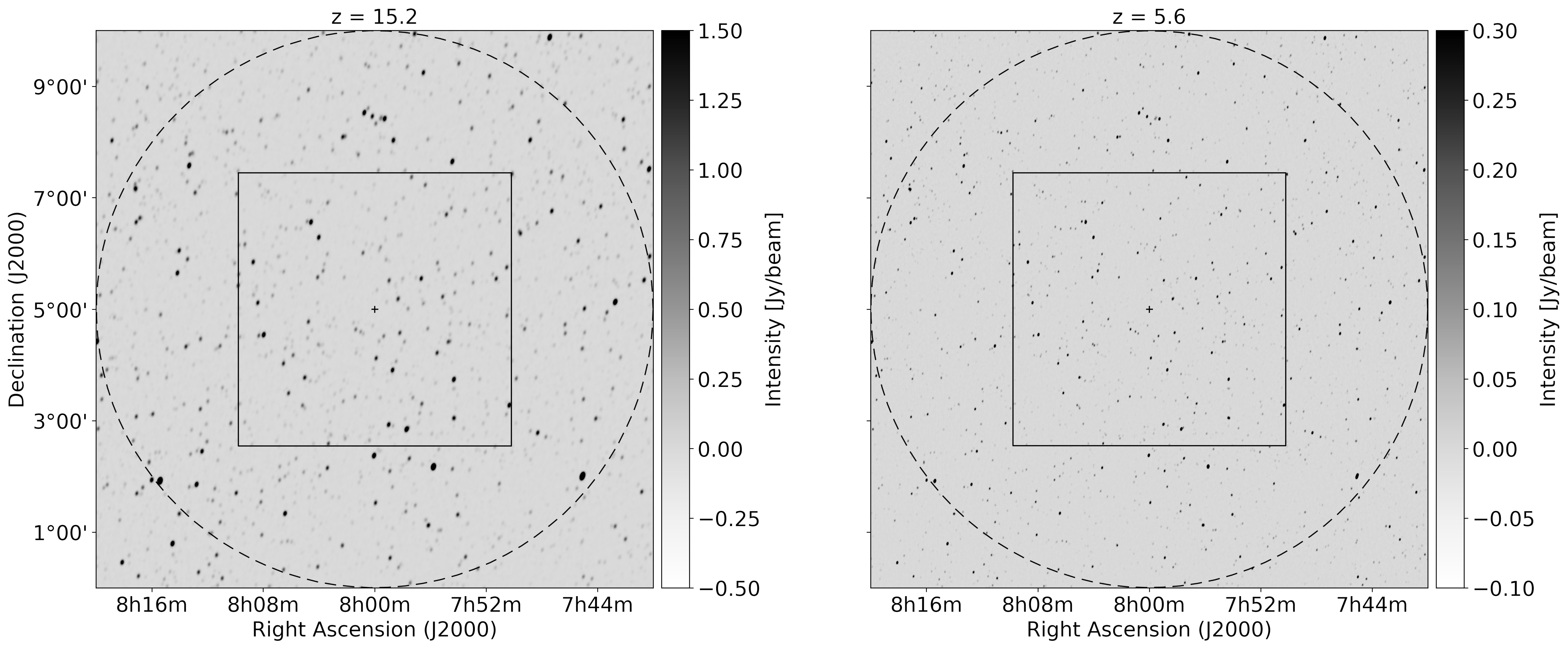}
 \caption{MWA G0044 field images. Left: combining 169 observation snapshots (338 minutes) at 87.675 MHz (z = 15.2); Right: combining 133 observation snapshots (266 minutes) at 215.675 MHz (z = 5.6), both covering regions of 100 degrees$^{2}$. The radius of dashed circle is 5 degrees, and the length of square is 4.9 degrees.}
\label{fig:org}
\end{center}
\end{figure*}

Figure \ref{fig:org} presents two deep wide-band images at 88 MHz and 216 MHz after calibration pipeline. Two images center at the same coordinate of [(RA, Dec) = (08:00:00, +05:00:00)], while both covering a region of 100 degrees$^{2}$ as shown in Figure \ref{fig:7fields}. The radius of dashed circle is 5 degrees, and the area it covers is used to analyze the statistical properties of radio sources in Section~\ref{sec:sources}. The length of the square is 4.9 degrees, and this region is utilized for foreground subtraction and power spectrum measurement in Section~\ref{sec:foregrounds}. The pixel size was set to 23.5 $\times$ 23.5 arcsec$^2$ at 88 MHz and 9.6 $\times$ 9.6 arcsec$^2$ at 216 MHz. 

Three groups of images are generated for studies with different purposes:
\begin{itemize}
    \item Combined images with a 30.72 MHz bandwidth are created to study the statistical properties of radio sources in low-frequency observations and to further establish the sky model for this region. 
    \item Images with a 30.72 MHz bandwidth but differing integration times are produced to assess the deep imaging quality for CD/EoR imaging. 
    \item Four sub-band images, each with a 7.68 MHz bandwidth, are generated for subsequent foreground subtraction and power spectrum measurement. 
\end{itemize}

As demonstrated in \citep{sokolowski2017calibration}, errors in the sky model or inaccuracies in the modelling of the MWA primary beam can introduce flux measurement errors of approximately 10\%, which are taken into account in the subsequent statistical measurements. We analysis the mosaiced synthesis beams at 88 MHz of integration time of 338 minutes and at 216 MHz of integration time of 266 minutes. The beam centers at [(RA, Dec) = (121.45, -3.97)] at 88 MHz while field of view (FOV) = 60 degrees, and [(RA, Dec) = (119.60, 1.11)] at 216 MHz, while FOV = 20 degrees. 

\section{Properties of Radio Sources in the Selected Field}
\label{sec:sources}

In this section, we describe the statistical properties of the radio sources. Source selection is performed on images with varying integration times, allowing us to compare the completeness of source identification across these different durations. The completeness of faint sources is achieved for the maximum integration time in our observations.

The purpose of considering different integration times is to determine where confusion becomes significant. The detection of faint sources can become limited by the confusion noise present in the image. In the absence of artifacts, the overall noise can be approximated as the sum of thermal noise and confusion noise. Thermal noise can be reduced by increasing the integration time, so comparing different integration times allows us to estimate the confusion limit and confusion noise.

\subsection{Source Finding}
\label{source_finding}

We analyse data from two frequency bands: 88 MHz and 216 MHz, considering different integration times individually. For the 88 MHz band, we use a max number of 169 snapshots (2 minutes observation time for each snapshot) in our observation. We evaluate the statistical properties of the sources using four integration times, which corresponding to stacking 5, 10, 40, and all 169 snapshots. Similarly, for the 216 MHz band, we used a maximum of 133 snapshots and considered four integration times based on stacking 5, 10, 40, and all 133 snapshots. 

In our study of the G0044 field, we investigate the statistical properties of radio sources in the circled sky region depicted in Figure \ref{fig:org}. We first run MIMAS tool from the AEGEAN \citep{hancock2012compact, hancock2018source} package to obtain a circular mask within the 5$^\circ$ radius region around the images center of [(RA, Dec) = (08:00:00, 05:00:00)]. Then, BAckground and Noise Estimation (BANE) tool from the AEGEAN package is run on the 30.72 MHz frequency bandwidth observations to derive the output position-dependent background and root-mean-squared (RMS) maps estimated for the images. Based on this information, we run AEGEAN to identify radio sources within the circular mask. The detected sources are characterised by AEGEAN as elliptical Gaussian components, with a source identification threshold set at five times of the local RMS (5$\sigma$). Notably, GLEAM catalogue employs the same 5$\sigma$ parameter for source detection \citep{hurley2017galactic}. Subsequently, we compare the number of radio sources detected in the G0044 field with the number of sources in the GLEAM catalogue. 

\subsection{Source Counts Analysis}

Here we investigate the source counts of the radio sources within the G0044 field. Source counts as a function of flux density are crucial for studying radio source populations and can offer a more realistic estimation of the anticipated foreground contamination in simulations for EoR experiments \citep{murray2017improved, nasirudin2020impact}. 

\subsubsection{Integrated Source Counts}

Our analysis of the G0044 field, specifically for the 216 MHz frequency band and with an integration time of 266 minutes, detects 2,576 radio sources. Importantly, all 971 sources in the same sky region listed in the GLEAM catalogue can be successfully matched with the sources detected in our analysis. As the integration time increases, the signal-to-noise ratio (SNR) of the images gradually improves, allowing more faint sources to be identified. Additionally, with the higher angular resolution, smaller-scale or combined radio sources may be resolved or distinguished in the G0044 field. 

\begin{figure*} 
\begin{center}
 \includegraphics[width=17.0cm]{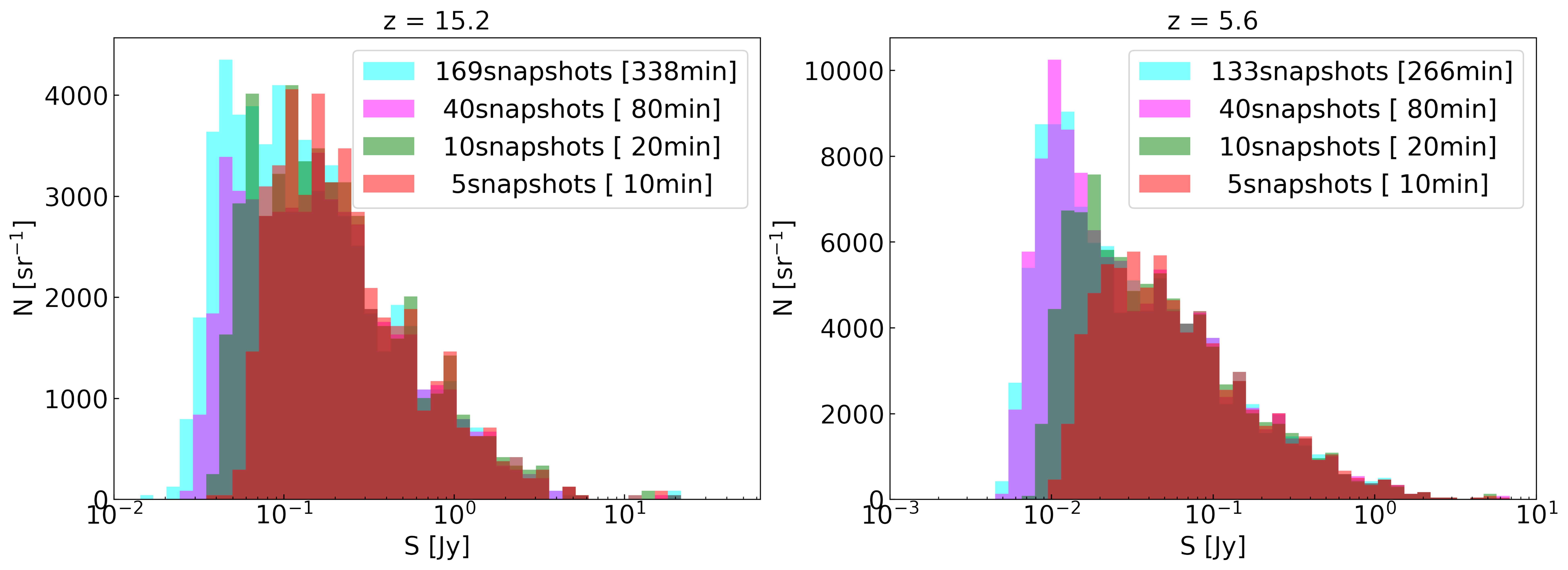}
 \caption{Integrated source counts ($N$) as a function of flux density ($S$). Left: 5 (red), 10 (green), 40 (magenta), 169 (cyan) observation snapshots results at 88 MHz; Right: 5 (red), 10 (green), 40 (magenta), 133 (cyan) observation snapshots results at 216 MHz.}
\label{fig:dN_87_215}
\end{center}
\end{figure*}

Figure \ref{fig:dN_87_215} presents the integrated source counts for our selected field, within a 5-degree radius, at two frequency bands and with different integration times. For flux densities above 83 mJy (88 MHz) and 22 mJy (216 MHz), the profiles align well across different integration times at high flux densities, indicating the detection of bright radio sources. However, at low flux densities, increasing the integration time reveals more faint radio sources. The lower detection at 88 MHz is attributed to the low angular resolution. The mean flux density and corresponding variance of the detected radio sources in the G0044 field are 0.33 Jy and 0.83 Jy at 88 MHz, and 0.097 Jy and 0.106 Jy at 216 MHz.

\subsubsection{Differential Source Counts}

Differential source counts d$N$/d$S$ in our selected sky region with different integration times are shown in the left panels of Figure \ref{fig:dNdS_E_87} at 88 MHz and Figure \ref{fig:dNdS_E_215} at 216 MHz. These data points are raw data without completeness-corrected. Uncertainties are represented by the Poisson errors calculated from the number of sources detected in each flux density bin, $\sigma_N = \sqrt{N}$.

We use the single power law fit for the sources with flux density above 49 mJy at 88 MHz and 10 mJy at 216 MHz, which can be expressed as:
\begin{equation}
 \frac{\rm{d} \textit{N}}{\rm{d} \textit{S}}=kS^{-\gamma} \rm Jy^{-1}\rm sr^{-1}.
\end{equation}
where $k$ and $\gamma$ are the fitting parameters. The best-fit parameters are: $k=2866 \pm 442, \gamma=1.96 \pm 0.10$ at 88 MHz with integration time of 338 minutes for flux density greater than 49 mJy, while $k= 2907 \pm 219, \gamma=1.67 \pm 0.03$ at 216 MHz with integration time of 266 minutes for flux density greater than 10 mJy. Differential source counts start to decrease for flux greater than 41 mJy at 88 MHz and 8.6 mJy at 216 MHz. 

\subsubsection{Euclidean-normalized Differential Source Counts}

\begin{figure*} 
\begin{center}
 \includegraphics[width=17.0cm]{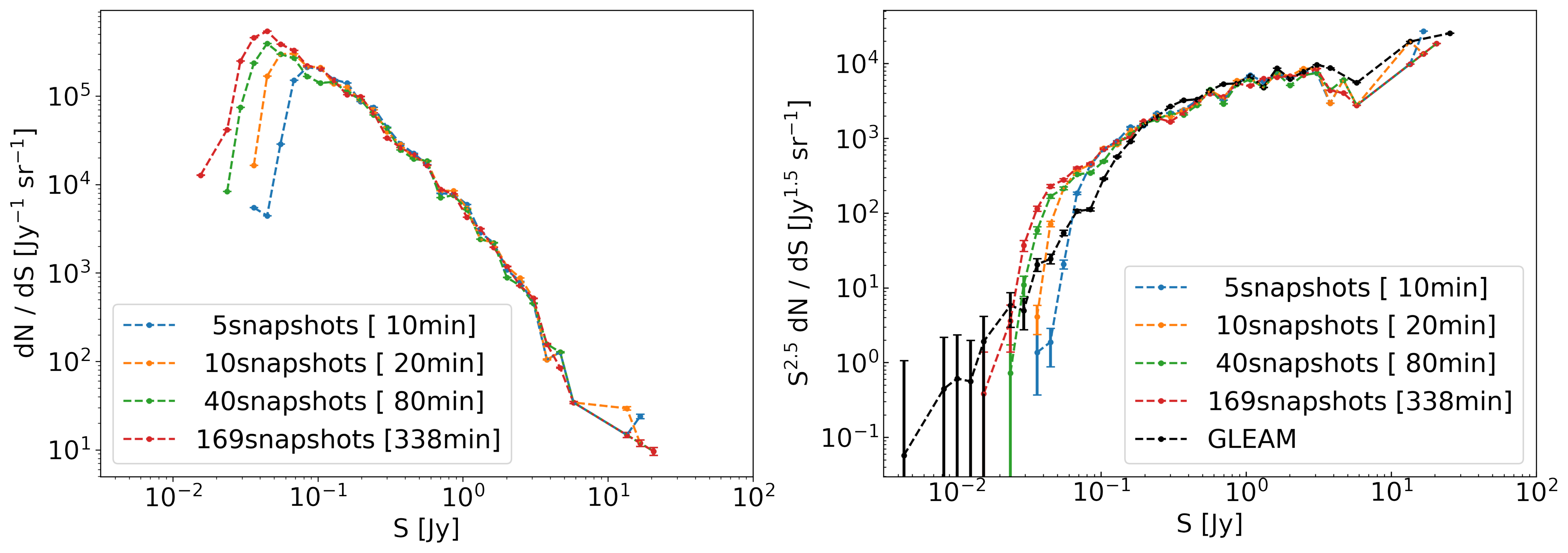}
 \caption{Differential source counts (d$N$ / d$S$) (left) and Euclidean-normalized differential source counts ($S^{2.5}$ d$N$ / d$S$) (right) as a function of flux density ($S$), with different integration time of 10 (cyan), 20 (magenta), 80 (green), 338 (red) minutes at 88 MHz. Euclidean-normalized differential source counts of GLEAM (black) at 84 MHz is shown for comparison.}
\label{fig:dNdS_E_87}
\end{center}
\end{figure*}

\begin{figure*} 
\begin{center}
 \includegraphics[width=17.0cm]{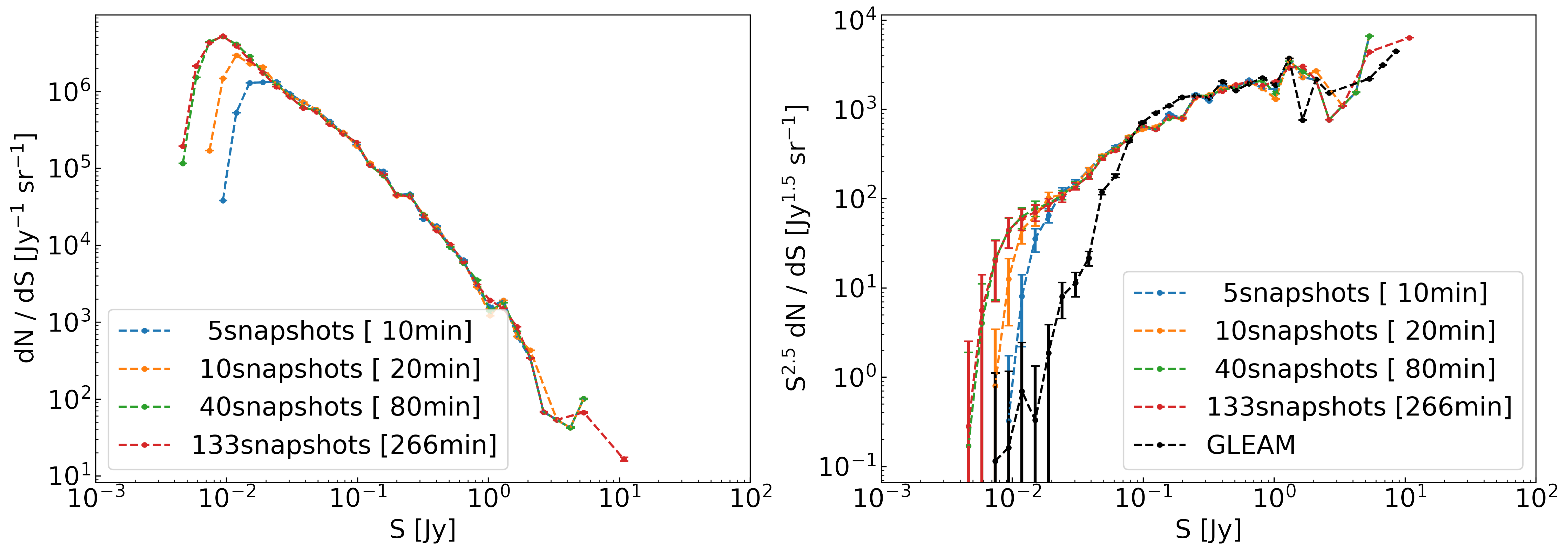}
 \caption{Differential source counts (d$N$ / d$S$) (left) and Euclidean-normalized differential source counts ($S{2.5}$ d$N$ / d$S$) (right) as a function of flux density ($S$), with different integration time of 10 (cyan), 20 (magenta), 80 (green), 266 (red) minutes at 216 MHz. Euclidean-normalized differential source counts of GLEAM (black) at 212 MHz is shown for comparison.}
\label{fig:dNdS_E_215}
\end{center}
\end{figure*}

We also derive the Euclidean-normalized differential source counts $S^{2.5}$d$N/$d$S$, and compare our analysis with GLEAM catalogue. Euclidean-normalized differential source counts of the detected radio sources with four integration times, and the results presented in GLEAM catalogue are shown in the right panels of Figure \ref{fig:dNdS_E_87} at 88 MHz and Figure \ref{fig:dNdS_E_215} at 216 MHz. Similarly, the uncertainties are represented by the Poisson errors calculated from the number of sources detected in each flux density bin. 

As the integration time increases, more sources with flux densities ranging from 37 mJy to 86 mJy at 88 MHz and from 10 mJy to 25 mJy at 216 MHz can be detected. With the longest integration time of 338 minutes at 88 MHz, even fainter sources with flux densities below 37 mJy down to 14 mJy can be detected. Similarly, at 216 MHz, sources with flux densities below 10 mJy, down to 5 mJy, are detectable with integration time of 266 minutes. Compared to the Euclidean-normalized differential source counts from the GLEAM catalogue, which has an integration time of approximately 3 hours per 2-minute observation with an average RMS of 2.7 mJy/beam at 154 MHz \citep{hurley2017galactic}, our study detects more faint sources within the same sky region, especially for those with flux densities below 130 mJy at 88 MHz and 79 mJy at 216 MHz.

Hence, unless specified otherwise, we will use the data with integration time of 338 minutes for the 88 MHz band and 266 minutes for the 216 MHz band in the subsequent calculations. 

\subsection{Noise Performance}
\subsubsection{RMS Noise}

\begin{figure*} 
\begin{center}
 \includegraphics[width=17.0cm]{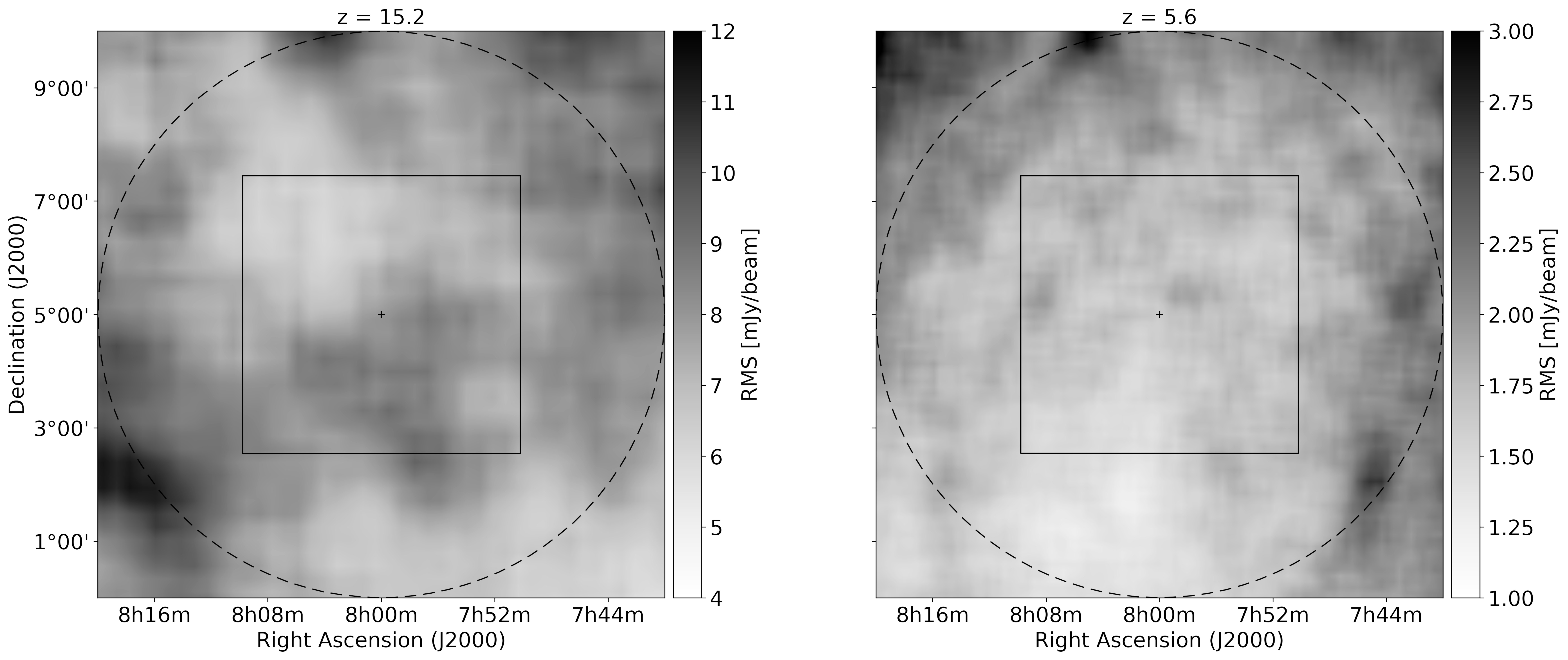}
 \caption{Root-Mean-Squared images. Left: 169 observation snapshots (338 minutes) at 87.675 MHz (z = 15.2); Right: 133 observation snapshots (266 minutes) at 215.675 MHz (z = 5.6), both covering region of 100 degrees$^{2}$. The radius of dashed circle is 5 degrees, and the length of square is 4.9 degrees.}
\label{fig:rms_88_216}
\end{center}
\end{figure*}

In order to estimate the sensitivity of the images, we use the root-mean-squared as an estimation of the image noise. We demonstrate the rms.fits generated from BANE, as shown in Figure \ref{fig:rms_88_216}, and then measure the average rms within the circular sky region. The results at two frequency bands with four different integration times are listed in Table \ref{tab:noise_estimates}. As the integration time increases, the RMS noise decreases, enabling the detection of fainter sources. The average RMS is 1.80 mJy/beam at 216 MHz with integration time of 266 minutes, which is comparable with the average RMS of 2.1 mJy/beam in \citet{lynch2021mwa}.

\begin{table*}
 \centering
 \caption{Measured and theoretical noise for different integration times at 88 and 216 MHz.}
 \label{tab:noise_estimates}
 \begin{tabular}{cc|cc|cc|cc}
  \hline
  \multicolumn{2}{c|}{Integration time {[}minutes{]}} & \multicolumn{2}{c|}{Measured RMS noise {[}mJy{]}} & \multicolumn{2}{c|}{Thermal noise {[}mJy{]}} & \multicolumn{2}{c}{Confusion noise {[}mJy{]}}\\
  at 88 MHz & at 216 MHz & at 88 MHz & at 216 MHz & at 88 MHz & at 216 MHz & at 88 MHz & at 216 MHz \\
  \hline
  10  & 10   & 14.98 & 3.74 & 16.79 & 2.63 & - & -  \\
  20  & 20   & 11.95 & 2.74 & 11.87 & 1.86 & - & -  \\
  80  & 80   &  9.00 & 1.93 & 5.93  & 0.93 & - & -  \\
  338 & 266  &  7.84 & 1.80 & 2.89  & 0.51 & 1.47 & 0.17 \\ 
  30000 & 30000 & - & - & 0.31 & 0.048 & - & -  \\ 
  \hline
 \end{tabular}
\end{table*}

\subsubsection{Thermal Noise}
We also give a prediction of the thermal noise level of the instrument, which reflects the sensitivity of the MWA and SKA1-Low, while the thermal noise can also be taken as the theoretical predicted RMS noise. The thermal noise of a radio interferometer $\sigma_{\rm N}$ with an effective area $A_{\rm eff}$ and the system temperature $T_{\rm sys}$ can be written as \citep{pal2024ionospheric}
\begin{equation}
\sigma_{\rm N} = \left( \frac{2k_{\rm B}}{A_{\rm eff} N_{\rm a}}\right) \left( \frac{T_{\rm sys}}{\sqrt{\Delta \nu t_{\rm int}}} \right) 
\end{equation}
where $k_{\rm B}$ is Boltzmann's constant, $N_{\rm a}$ is the number of antennas, $\Delta \nu$ is the channel width, $t_{\rm int}$ is the integration time. The system temperature $T_{\rm sys}$ is composed of the sky temperature $T_{\rm sky}$ and the receiver temperature $T_{\rm rec}$, that is $T_{\rm sys} = T_{\rm sky} + T_{\rm rec}$. The sky temperature $T_{\rm sky}$ can be described approximately as $T_{\rm sky} = 60\rm K(\nu/300 \rm MHz)^{-2.25}$ \citep{tingay2013murchison} and the receiver temperature $T_{\rm rec} = 28 \rm K$ \citep{wayth2018phase}. 

Considering MWA instrument of Phase II extended configuration, the channel width $\Delta \nu = 0.75 \times$ 7.68 MHz given a 25\% reduction in the channel width due to flagged edge channels, $A_{\rm eff} = 21.5\rm m^2$, $N_{\rm a} = 128$. So the thermal noise of MWA at 88 MHz with a system temperature of 983 K is 2.89 mJy and 0.31 mJy for integration time of 338 minutes and 500 hours. And the thermal noise of MWA at 216 MHz with a system temperature of 154 K is 0.51 mJy and 0.048 mJy for integration time of 266 minutes and 500 hours. 
We also estimate the thermal noise level for SKA1-Low. For SKA1-Low, the channel width $\Delta \nu$ = 125 kHz, $N_{\rm a} = 512$, $A_{\rm eff} / T_{\rm sys}$ is supposed to be 399.7 $\rm m^2 / K$ at 88 MHz, and 646.5 $\rm m^2 / K$ at 216 MHz. For integration time of 500 hours, the thermal noise of SKA1-Low is 0.028 $\mu$Jy and 0.018 $\mu$Jy at 88 MHz and 216 MHz. The details of the parameters used in MWA are listed in Table \ref{tab:MWA_parameters} and SKA1-Low in Table \ref{tab:SKA_parameters}.

\subsubsection{Confusion Noise}

We adopt the differential source counts d$N/$d$S$ derived from the G0044 field to predict the confusion noise for the MWA and future SKA1-Low observation. Confusion noise $\sigma_{\rm c}$ arising from the uncertainties in the position of unresolved faint radio sources below a flux threshold $S_{\rm lim}$ can be estimated through \citet{scheuer1957statistical, condon1974confusion, condon2012resolving}: 
\begin{equation}
\sigma_{\rm c}^2=\Omega_{\rm b} \int_{0}^{\textit{S}_{\rm{lim}}}\frac{\rm{d}\textit{N}}{\rm{d}\textit{S}} \textit{S}^2 \rm{d}\textit{S}
\end{equation}
where $\Omega_{\rm b}$ is the finite synthesis beam of the radio interferometer, 
$\Omega_{\rm b} = \frac{\pi \theta_{\rm s}^2}{4\ln{2}}$ where the beam size $\theta_{\rm s} = 1.15 \rm arcmin (\nu / 154 \rm MHz) ^{-1}$ for MWA Phase II. $S_{\rm lim}$ is the flux threshold at which the source density becomes too high for the array to clearly resolve the sources. This threshold is determined by the errors of calibration, deconvolution, and the resolution. Specifically, $S_{\rm lim} = 5\sigma$, where $\sigma$ represents the RMS noise.

Considering MWA instrument of Phase II extended configuration, we adopt $\theta_{\rm s}$ to 2.02 arcmin at 88 MHz and 0.82 arcmin at 216 MHz. $S_{\rm lim}$ is set to 39.2 mJy with 73.55\% completeness level at 88 MHz with integration time of 338 minutes, and 9 mJy with 83.39\% completeness level at 216 MHz with integration time of 266 minutes. So the confusion noise of MWA is 1.47 mJy at 88MHz and 0.17 mJy at 216 MHz. 
Considering SKA1-Low, we adopt $\theta_{\rm s}$ to 17 arcsec at 88 MHz and 6.4 arcsec at 216 MHz. So the confusion noise of SKA1-Low is 0.21 mJy at 88 MHz for the 338 minutes deep imaging of the 'G0044' field and 0.022 mJy at 216 MHz for the 266 minutes deep imaging. 

Based on our theoretical prediction, the confusion noise is higher than the thermal noise for future SKA1-Low observation. For future deep imaging of EoR using SKA1-Low, confusion noise will be the primary factor to consider in detecting faint sources, particularly in a sky field similar to the G0044 field. Although, it should be noted that when calculating the confusion noise of SKA1-Low, we use the source count derived from the MWA observation. In addition, with future SKA1-Low instrument, we anticipate improved sensitivity and resolution, along with reduced calibration errors, leading to a lower confusion noise compared to our current estimations.

\begin{table}
\centering
\caption{Parameters of the MWA observations}
\label{tab:MWA_parameters}
\begin{tabular}{lc}
\hline
    Parameter & Value\\
    \hline
    UTC Start & \makecell[c]{2018-04-13T10:20:06.000Z -- \\ 2018-04-30T11:33:18.000Z}\\
    MWA Configuration & Phase II Extended \\
    Central Frequency  & \makecell[c]{88 MHz (z $\sim$ 15.20) \\ 216 MHz (z $\sim$ 5.58)}\\
    Bandwidth & 30.72 MHz\\
    Spectral Resolution & 40 kHz\\
    Number of Antenna Tiles ($N_{\rm a}$) & 128\\
    Max Number of Used Snapshots & \makecell[c]{169 at 88 MHz \\ 133 at 216 MHz}\\
    Integration Time per Snapshots & 2 minutes\\
    Imaging Center (J2000) & (RA, Dec) = (8:00:00, +5:00:00)\\
    Effective Collective Area ($A_{\rm eff}$) & 21.5 $\rm m^2$ \\
    System temperature ($T_{\rm sys}$) & \makecell[c]{983 K at 88 MHz \\ 154 K at 216 MHz} \\
    Beam size ($\theta_{\rm s}$) & \makecell[c]{2.02 arcmin at 88 MHz \\ 0.82 arcmin at 216 MHz} \\
\hline
\end{tabular}
\end{table}

\begin{table}
\centering
\caption{Parameter estimation of the future SKA1-Low}
\label{tab:SKA_parameters}
\begin{tabular}{lc}
\hline
    Parameter & Value\\
    \hline
    Central Frequency  & \makecell[l]{88 MHz (z $\sim$ 15.20) \\ 216 MHz (z $\sim$ 5.58)}\\
    Spectral Resolution & 125 kHz\\
    Number of Array Elements ($N_{\rm a}$) & 512 \\
    $A_{\rm eff} / T_{\rm sys}$ & \makecell[c]{399.7 $\rm m^2 / K$ at 88 MHz \\ 646.5 $\rm m^2 / K$ at 216 MHz} \\
    Beam size ($\theta_{\rm s}$) & \makecell[c]{17 arcsec at 88 MHz \\ 6.4 arcsec at 216 MHz} \\
    \hline
\end{tabular}
\end{table}

\subsection{Source Catalogue}

The final catalogue is derived from the total 2,576 sources detected at 216 MHz, offering a better angular resolution compared to the 88 MHz image. For each source, we list properties as reported by AEGEAN, such as Right Ascension (J2000) and Declination (J2000). Additionally, we estimate the fitted spectral index and its associated error using the methodology outlined in \citet{he2024influence}, which fitting data from GLEAM, NVSS, SUMSS, the Rapid ASKAP Continuum Survey (RACS; \citet{hale2021rapid}) and our two frequency bands. Notably, the spectral index of SP sources in \citet{he2024influence} is marked as "inf". The sources are ranked based on their integrated flux density at 216 MHz. Table \ref{tab:source_cat} presents the top 20 sources, while Table \ref{tab:column_name} provides a detailed description of the corresponding columns.

We also list the number of sources at 216 MHz unmatched with GLEAM and their contributions to the integrated flux with flux range in Table \ref{tab:dif_S-N}, and the number of sources at 216 MHz and their contributions to the integrated flux with flux cut in Table \ref{tab:S216_N}.

\begin{table}
\centering
\caption{Number of sources at 216 MHz unmatched with GLEAM and their contributions to the integrated flux.}
\label{tab:dif_S-N}
\begin{tabular}{c|cc}
\hline
$S$$_{\rm 216MHz}$ {[}Jy{]} & $N$$_{\Delta}$ & Fraction of total flux (\%) \\
\hline
10$^{-1}$ - 10$^{0}$    & 10       & 0.80    \\
10$^{-2}$ - 10$^{-1}$   & 1126     & 9.86    \\
10$^{-3}$ - 10$^{-2}$   & 469      & 1.51    \\
\hline
\end{tabular}
\end{table}

\begin{table}
\centering
\caption{Number of sources at 216 MHz and their contributions to the integrated flux.}
\label{tab:S216_N}
\begin{tabular}{c|ccc}
\hline
\begin{tabular}[c]{@{}c@{}}Flux cut \\ $S$$_{\rm 216MHz}$ {[}Jy{]}\end{tabular} & \begin{tabular}[c]{@{}c@{}}Number \\ of sources\end{tabular} & \begin{tabular}[c]{@{}c@{}}Fraction of \\ total flux (\%)\end{tabular} & \begin{tabular}[c]{@{}c@{}}Fraction of \\ total number(\%)\end{tabular} \\
\hline
10$^{0}$             & 33           & 26.90         & 1.28        \\
10$^{-1}$            & 486          & 76.46         & 18.87       \\
10$^{-2}$            & 2107         & 98.49         & 81.79       \\
10$^{-3}$ (No cut)   & 2576         & 100           & 100         \\ 
\hline
\end{tabular}
\end{table}

\section{Catalogue Properties}
\label{sec:catalogue}
Following the procedure outlined by \citet{lynch2021mwa} for the MWA Long Baseline Epoch of Reionisation Survey (LoBES), we conduct analysis encompassing flux error assessment, position offset evaluation to gauge data quality, 
examination of spectral indices distribution and source classification, and implementation of additional catalogue corrections.

\subsection{Error Estimation}

\begin{figure*}
\centering 
 \includegraphics[width=17.0cm]{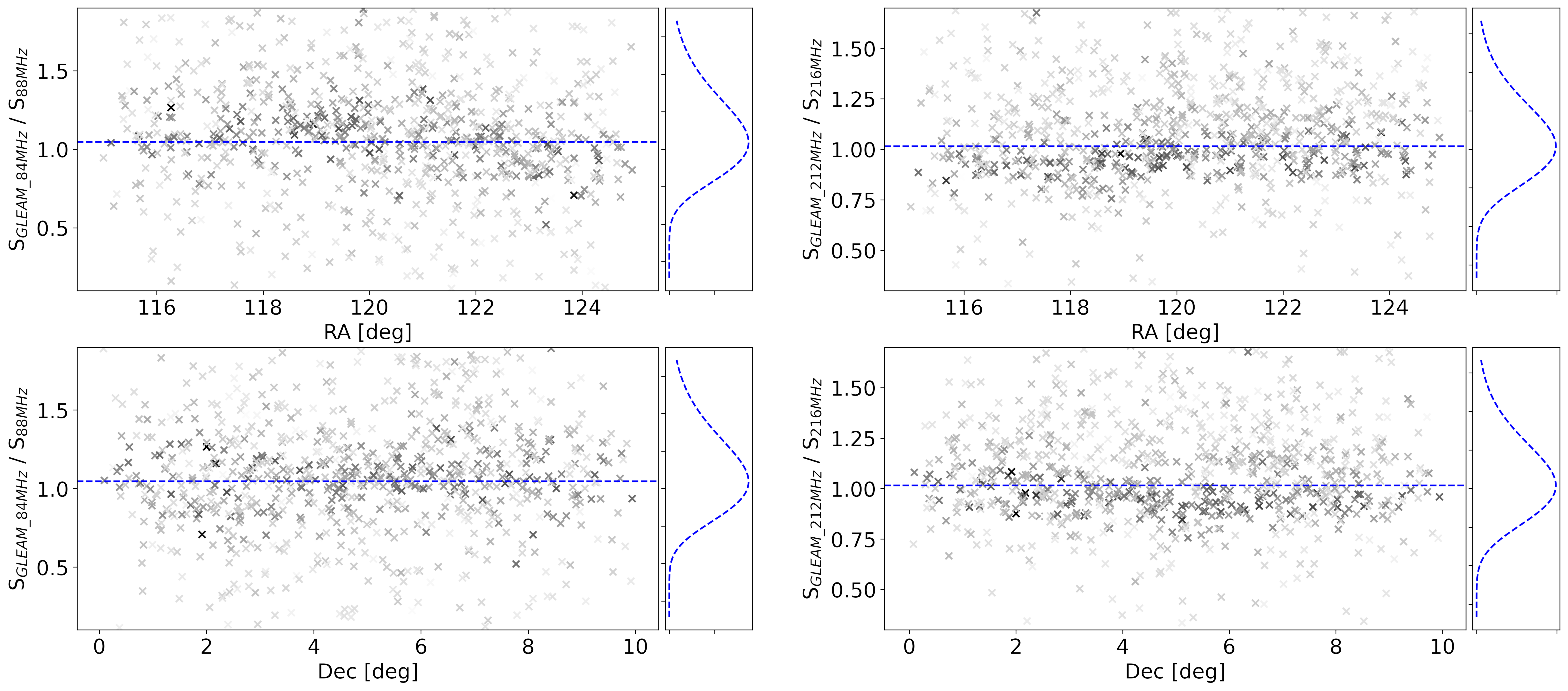}
 \caption{The distribution between the flux ratio and RA (top row) or Dec (bottom row), where the ratio is the flux density ratio of all matched sources in (left column) GLEAM at 84 MHz and our 88 MHz results; (right column) GLEAM at 212 MHz and our 216 MHz results.}
\label{fig:flux_ratio_ra_dec_88_216}
\end{figure*}

We plot the distribution between the flux ratio and RA (top row) or Dec (bottom row) in Figure \ref{fig:flux_ratio_ra_dec_88_216}, where the ratio is the flux density ratio of all matched sources in (left column) GLEAM at 84 MHz and our 88 MHz results; (right column) GLEAM at 212 MHz and our 216 MHz results. Log-normalized SNR has been used as the gray color for each point in each subgraph. We also plot the fitted log-Gaussian weighted curve at the right panel of each subgraph, using the SNR as the weight, and the dashed horizontal line is the median value of the fitted curve. We find a mean ratio of 1.05 and 1.02, with a standard deviation of 0.12 and 0.09, at 88 MHz and 216 MHz. Generally for sources with high-SNR (dark gray), the flux density in GLEAM is close to our detected results.

\begin{figure}
\centering 
 \includegraphics[width=8.0cm]{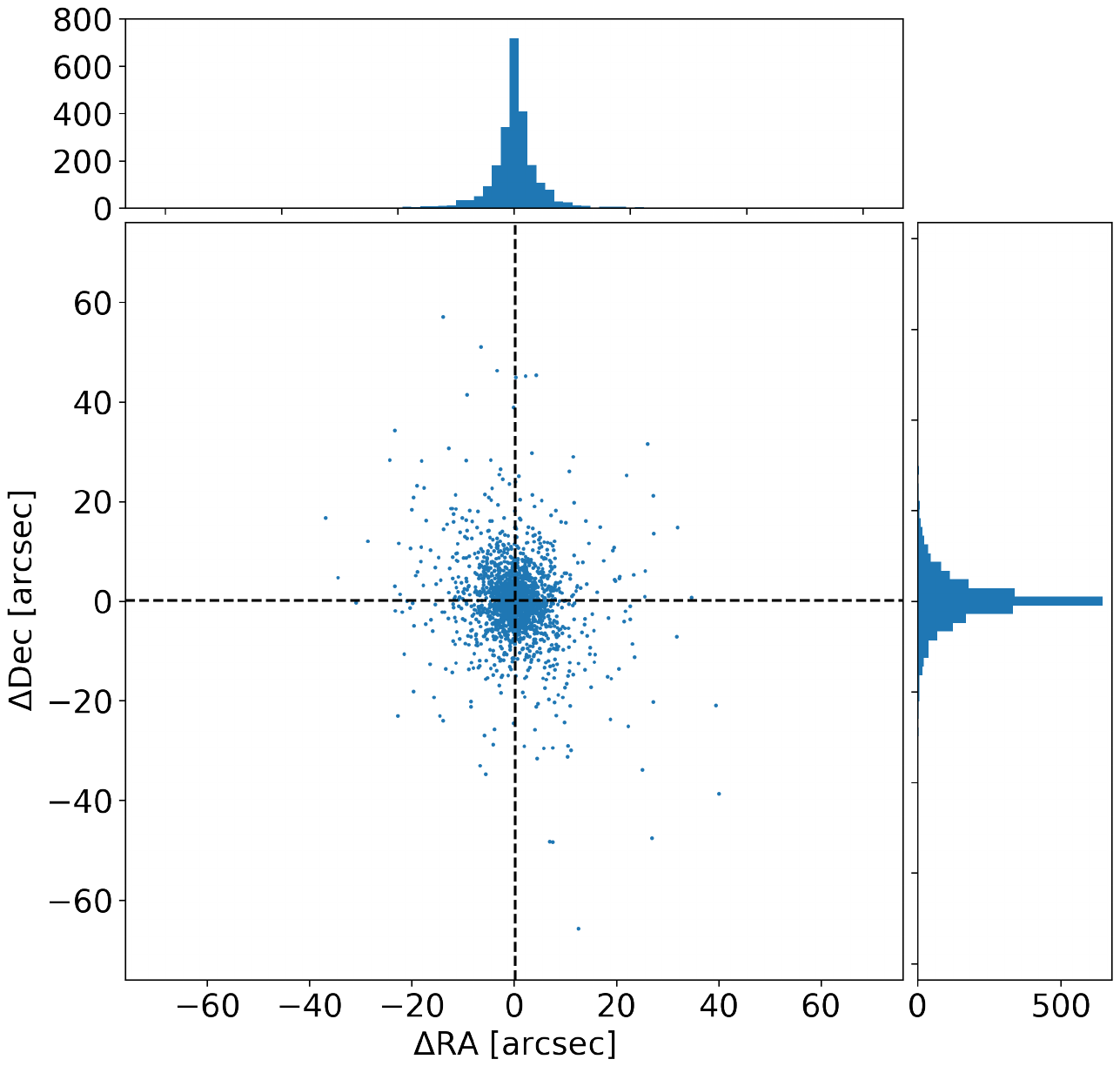}
 \caption{Source position offsets in RA and Dec when matching with NVSS. We also show the histograms of the differences in RA (top) and Dec (right), and the dashed lines are the mean value of these differences, which are closing to zero.}
\label{fig:position_offset_NVSS_216}
\end{figure}

Apart from the flux density error estimation, we also analysis the source position offset. Figure \ref{fig:position_offset_NVSS_216} shows the distribution of differences in RA and Dec for 2388 sources matched with NVSS catalogue. Dashed lines are the mean value of source position offset, that $\Delta$ RA = 0.19 arcsec and $\Delta$ Dec = 0.15 arcsec are closing to zero. Considering the small scatter in these offsets, we do not make any corrections for the offsets. 

\subsection{Spectral Indices}
\begin{figure}
\centering
 \includegraphics[width=8.0cm]{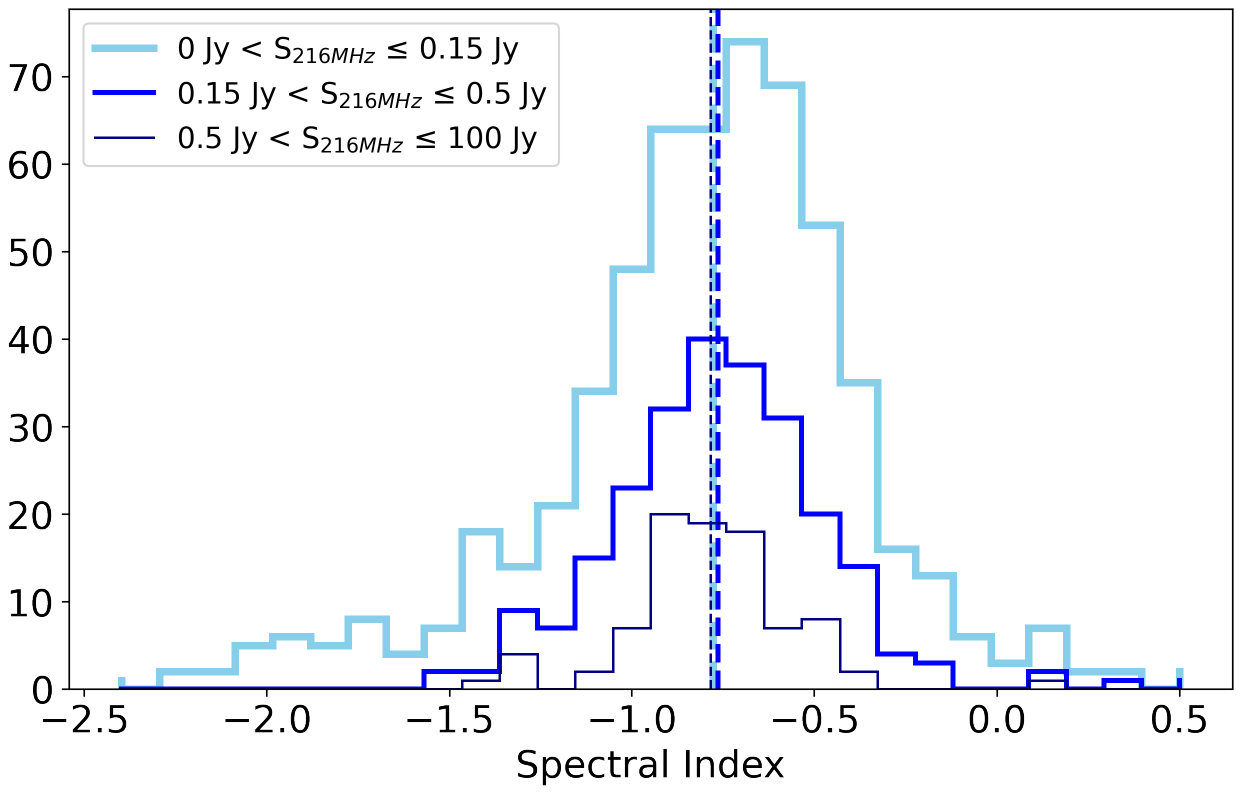}
 \caption{Spectral index distribution for sources that are well-fitted by the method presented in \citet{he2024influence} is analyzed. The median values within flux density ranges at 216 MHz for 0 to 0.15 Jy, 0.15 to 0.5 Jy and 0.5 to 100 Jy are -0.78, -0.76 and -0.78.}
\label{fig:alpha_216}
\end{figure}

Given the well-fitted spectral indices obtained through the method presented in \citet{he2024influence}, which fitting data from GLEAM, NVSS, SUMSS, RACS and our two frequency bands, we analyze its distribution (also recorded in Table \ref{tab:source_cat}) across different flux density ranges at 216 MHz for 0 to 0.15 Jy (596 sources), 0.15 to 0.5 Jy (242 sources) and 0.5 to 100 Jy (89 sources), as shown in Figure \ref{fig:alpha_216}. Correspondingly, the median spectral index is -0.78 $\pm$ 0.02, -0.76 $\pm$ 0.02 and -0.78 $\pm$ 0.02, for each flux density range. Several other papers have reported spectral indices for sources identified within the low-frequency range, and our fitted spectral indices are generally in agreement. Specifically, the LoBES reports spectral indices of -0.75 $\pm$ 0.16 for sources with flux densities ranging from 50 to 150 mJy at 204 MHz, while the GLEAM survey reports spectral indices of -0.79 for sources with flux densities above 60 mJy at 200 MHz.

\subsection{Classification of Sources}

As approach outlined in \citep{franzen2015atlas, franzen2019source} we identify extended sources within our 5-degree radius using the ratio of the integrated flux density, $S_{\rm int}$, to the peak flux density, $S_{\rm peak}$, for each source identified by AEGEAN. To detect source extension at the 2$\sigma$ level, we follow
\begin{equation}
\ln{\left ( \frac{S_{\rm int}}{S_{\rm peak}} \right )} > 2 \sqrt{\left (\frac{\sigma_{\rm int}}{S_{\rm int}}\right )^2 + \left (\frac{\sigma_{\rm peak}}{S_{\rm peak}}\right )^2}
\end{equation}
where $\sigma_{\rm int}$ and $\sigma_{\rm peak}$ represent the uncertainties in the integrated and peak flux densities, respectively, and are the Gaussian parameter fitting uncertainties returned by AEGEAN. 

\begin{figure*} 
\begin{center}
 \includegraphics[width=17.0cm]{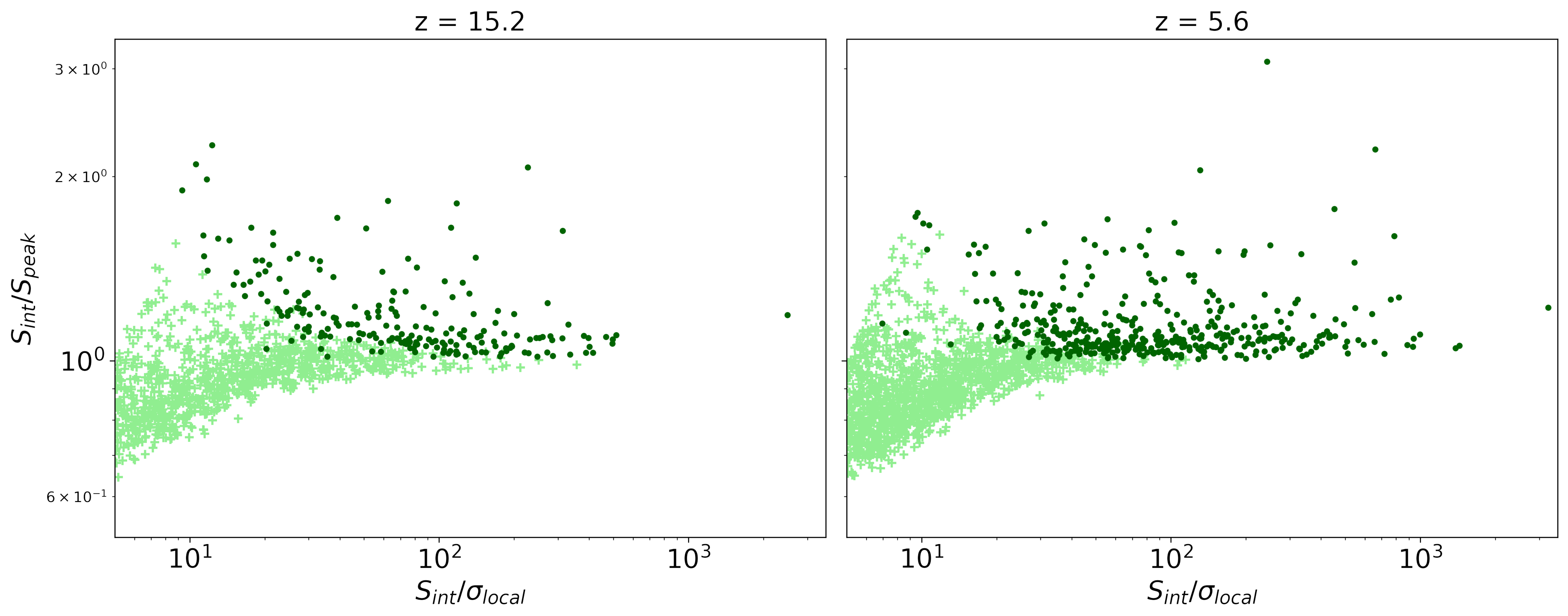}
 \caption{Ratios of the integrated to peak intensity as a function of the signal-to-noise ratio of the detected source within 5-degree radius. The dark green circles represent extended sources while the light green crosses are point-like sources. Extended sources account for 15.0\% at 88 MHz and 16.5\% at 216 MHz.}
\label{fig:class_88_216}
\end{center}
\end{figure*}

Figure \ref{fig:class_88_216} shows the ratio of the integrated to the peak flux densities as a function of the signal-to-noise ratio for the sources detected at 88 MHz (left panel) and 216 MHz (right panel). Extended sources are represented by dark green circles, while point-like sources are shown as light green crosses. Among the detected sources, 15.0\% and 16.5\% are classified as extended at 88 MHz and 216 MHz, while 22\% of detected sources in the LoBES wide-band image at 200 MHz are classified as extended. We also notice that there are fewer extended sources compared to point-like sources at low SNR in our selected sky region.

\subsection{Additional Corrections}

The likelihood of detecting sources of varying sizes and flux densities in our catalogue is influenced by their position within the images and their SNR. Consequently, the source counts obtained from AEGEAN do not accurately reflect the true extragalactic source count distribution. To calculate the expected extragalactic source counts, it is necessary to compute and apply several correction factors to the source counts generated by our source finding procedure. We will outline these correction factors in the following sections.

\subsubsection{False Detection Rate}

To address noise spikes and artefacts in the images that AEGEAN flagged as real radio sources, we utilize the false detection rate (FDR). Given the symmetrical nature of the image noise, we anticipate that positive noise spikes will have counterpart negative spikes. By analysing the inverted (negative) images, we can detect these negative spikes. To accomplish this, we run AEGEAN using the same parameters employed during the creation of the source catalogue. 

We only identify 2 sources in inverted images at 88 MHz with integration time of 338 minutes and at 216 MHz with integration time of 266 minutes. Both of these 2 sources are around bright sources, and \citet{lynch2021mwa} has excluded regions around bright sources with flux densities greater than 5 Jy. Therefore, we have accepted all the sources listed in the catalogue without any exclusion.

\subsubsection{Source Completeness}

We also conduct simulations to evaluate the completeness of our source selection individually across two frequency bands, which are also performed in \citet{lynch2021mwa}. We quantify the various biases in our catalogue by injecting 500 simulated sources into 88 MHz image and 1000 simulated sources into 216 MHz image. These sources are randomly distributed within our 5-degree radius region, with a minimum enforced distance of 5 arcminutes between any two simulated sources. We assign a random source size and position angle to each simulated source, within the range of the source parameters generated by AEGEAN in the two frequency-band images, respectively. The flux densities of the simulated sources were drawn randomly from the power-law distribution of d$N/$d$S \propto S^{-1.59}$ between 5 mJy and 10 Jy at 153 MHz \citep{intema2011deep, williams2013t}, and then scaled to the 88 / 216 MHz assuming the spectral index of -0.8. In order to generate good statistical data, we run 20 such simulations.

To inject these simulated sources into the images we utilize AERES from the AEGEAN package. For each simulation we then execute the same source finding procedures outlined in Section~\ref{source_finding}. We now have a collection of 20 source catalogues, each containing both the detected injected sources and the original image sources. After cross-matching between each of the 20 catalogues with the original simulated source catalogue, we may identify which of the simulated sources have been recovered in each simulations. To account for source confusion between the injected sources and the original sources in two frequency-band images, we then cross-match the detected sources with the original image sources. We remove the matched sources that have fitted source positions closer to a matched original image source than to a matched injected source.

The correction factor for the $i$-th flux density bin is then computed as follows:
\begin{equation}
 \rm{Completeness \ Correction}_\textit{i} = \frac{\textit{N}_{\rm{simulated}, \textit{i}}}{\textit{N}_{\rm{recovered}, \textit{i}}}
\end{equation}
The final correction factor for each flux density bin is determined by taking the median of the correction values obtained from the 20 simulations, while the uncertainties are calculated as the 16th and 84th percentile values, as are listed in Table \ref{tab:corrected_endc}. In the table, we notice that the completeness correction contributes more to the low flux density bins. The completeness of the catalogue at a given flux density is then calculated from integrating the detected fraction of radio sources (that is the inverse of the completeness correction value) upwards from a given flux density threshold. We estimate that the catalogue is 90\% complete above 81 mJy at 88 MHz and 10.4 mJy at 216 MHz, as are listed in Table \ref{tab:catalogue_properties}.

\begin{table}
	\centering
	\caption{The catalogue properties.}
	\label{tab:catalogue_properties}
	\begin{tabular}{l|cc}
	\hline
    Central frequency [MHz]& 87.675 & 215.675 \\
    Field of view [deg$^2$]& \multicolumn{2}{c}{78.54}     \\
    Pixel size [arcsec/pixel]& 23.5 $\times$ 23.5 & 9.6 $\times$ 9.6  \\
    rms (mean) [mJy/beam]  &  7.84 & 1.80 \\
    Number of sources      & 1,390   & 2,576    \\
    Number of matched sources & \multicolumn{2}{c}{1,381} \\
    80\% completeness [mJy]& 45.8   & 7.7     \\
    90\% completeness [mJy]& 81.0   & 10.4    \\
    \hline
	\end{tabular}
\end{table}

\subsubsection{Final Source Counts}

To calculate the corrected source counts in our selected field at 216 MHz, we multiply the source counts in each flux density bin by the corresponding completeness corrections factor. We have not considered FDR here since no sources are excluded in this work. The uncertainties of the corrected source counts are from the completeness correction. To facilitate comparison with other results, we scale the flux density bins from 216 MHz to 154 MHz assuming a spectral index of -0.8, while the values in each scaled flux density bin are listed in Table \ref{tab:corrected_endc}. When compared to the findings reported in \citet{lynch2021mwa}, our corrected source counts are lower in the comparable flux density bins. This discrepancy may suggest that our selected sky region is relatively ‘quiet’ in terms of source detection.

\begin{table*}
\centering
\caption{The Euclidean-normalised differential source counts for the G0044 field scaled from 216 MHz to 154 MHz. The range of the flux density bin is given by $S_{\rm range}$, with a center flux density of $S_{\rm c}$. $N$ is the total number of uncorrected sources per bin and the last column gives the corrected normalised source counts ($S$$^{2.5}$ d$N$/d$S$).}
\label{tab:corrected_endc}
\begin{tabular}{ccccc}
\hline
$S_{\rm range}$ & $S_{\rm c}$  & $N$ & Completeness Correction & Corrected $S^{2.5} $d$N/$d$S$ \\
$[\rm{Jy}]$ & $[\rm{Jy}]$ & & & $[\rm{Jy}^{3/2}$ sr$^{-1}]$ \\
\hline
0.006 - 0.010  & 0.008    & 142 & 1.486 $\pm$ 0.027 & 15 $\pm$ 1    \\
0.010 - 0.014  & 0.012    & 430 & 1.199 $\pm$ 0.015 & 71 $\pm$ 1    \\
0.014 - 0.022  & 0.018    & 417 & 1.062 $\pm$ 0.011 & 111 $\pm$ 1   \\
0.022 - 0.032  & 0.027    & 303 & 1.039 $\pm$ 0.010 & 145 $\pm$ 1   \\
0.032 - 0.048  & 0.040    & 238 & 1.028 $\pm$ 0.008 & 207 $\pm$ 1   \\
0.048 - 0.072  & 0.060    & 239 & 1.020 $\pm$ 0.008 & 378 $\pm$ 2   \\
0.072 - 0.108  & 0.090    & 218 & 1.015 $\pm$ 0.010 & 630 $\pm$ 6   \\
0.108 - 0.162  & 0.135    & 182 & 1.013 $\pm$ 0.010 & 962 $\pm$ 9   \\
0.162 - 0.243  & 0.202    & 128 & 1.009 $\pm$ 0.009 & 1234 $\pm$ 10 \\
0.243 - 0.363  & 0.303    & 94  & 1.011 $\pm$ 0.011 & 1665 $\pm$ 18 \\
0.363 - 0.544  & 0.454    & 71  & —— & 2278  \\
0.544 - 0.815  & 0.679    & 51  & —— & 2999  \\
0.815 - 1.220  & 1.017    & 26  & —— & 2801  \\
1.220 - 1.827  & 1.523    & 19  & —— & 3750  \\
1.827 - 2.735  & 2.281    & 13  & —— & 4702  \\
2.735 - 4.095  & 3.415    & 1   & —— & 662   \\
4.095 - 6.132  & 5.113    & 1   & —— & 1214  \\
6.132 - 9.181  & 7.656    & 2   & —— & 4449  \\
9.181 - 13.74  & 11.46    & 1   & —— & 4076  \\ 
\hline                           
\end{tabular}
\end{table*}

\section{Foregrounds Removal}
\label{sec:foregrounds}
Foreground contaminants are significantly brighter than the cosmological signal by several orders of magnitude, posing a major challenge for the analysis of upcoming observational data to separate the two signals \citep{de2008model, di2002radio}. In this section, we present the foreground removal algorithm employed in this work and demonstrate its application through statistical power spectrum measurements. Our pipeline accounts for direction-dependent (DD) effects (e.g., ionospheric distortions and primary beam variations) during post-processing corrections. Instead of subtracting sources during the calibration process \citep{mertens2020improved}, we employ PCA for foreground removal.

\subsection{Foregrounds Removal Algorithm}

Fortunately, foregrounds are expected to be spectrally smooth across frequency \citep{liu2012well, tegmark2000foregrounds, adam2016planck}, whereas the cosmological signal is expected to vary with frequency \citep{liu2020data, di2002radio}. Current separation techniques thus rely on the statistical distinctions among 21cm spectral components. The observed 21cm signal can be modelled as the sum of three components: foreground modes, cosmological signals and noise, written as: 
\begin{equation}
T_{\rm obs}(\nu, \hat{\textbf{n}}) = T_{\rm fg}(\nu, \hat{\textbf{n}}) + T_{\rm cosmo}(\nu, \hat{\textbf{n}}) + T_{\rm noise}(\nu)
\end{equation}
where each component is described with respect to a given frequency, $\nu$, and line of sight direction, $\hat{\textbf{n}}$. Moreover, the foreground modes can be modelled as the sum of four components: galactic synchrotron emission, galactic free-free emission, point sources and radio halos, written as: 
\begin{equation}
T_{\rm fg}(\nu, \hat{\textbf{n}}) = T_{\rm syn}(\nu, \hat{\textbf{n}}) + T_{\rm free}(\nu, \hat{\textbf{n}}) + T_{\rm point}(\nu, \hat{\textbf{n}}) + T_{\rm halo}(\nu, \hat{\textbf{n}})
\end{equation}

Principal Component Analysis (PCA) makes use of the statistical properties of foreground signals by fitting both foreground sky maps and foreground functions simultaneously \citep{makinen2021deep21, alonso2015blind, de2008model}. Conceptually, PCA can be considered as fitting a multidimensional ellipsoid to a feature space, with eigenvectors pointing to the largest variance. Since foregrounds are expected to exhibit smoothness and strong frequency correlation \citep{de2008model, di2002radio, pritchard201221}, eliminating the components associated with largest eigenvalue \citep{alonso2015blind} is expected to retain the cosmological signal on large angular scales. The method has been used for foregrounds removal in both simulated and real 21cm data \citep{chang2010hydrogen, masui2013measurement, switzer2015interpreting}.

We use four sub-band images of the G0044 field, created by using WSClean, which split the total 30.72 MHz bandwidth into four 7.68 MHz channels. These sub-band images serve as the basis for statistical measurements and are prepared for subsequent foreground removal in the observation temperature map. The FOV of the images is 4.9 $\times$ 4.9 degrees$^2$ around the center of the images [(RA, Dec) = (8:00:00, +5:00:00)] as shown in the square of Figure \ref{fig:org}. After reading the beam parameters from the deep images, we calculate the average beam size among each sub-band images, through $\rm{beam}_{major} \times \rm{beam}_{minor} \frac{\pi}{4\ln{2}}$. Then, we convert the image unit from [Jy/beam] to [K] to obtain the observation temperature map, taking into account the telescope's beam size using the formula $T_{\rm B} \equiv S_\nu / (2k\nu^2/{\rm c}^2)$.

\begin{figure*} 
\begin{center}
 \includegraphics[width=17.0cm]{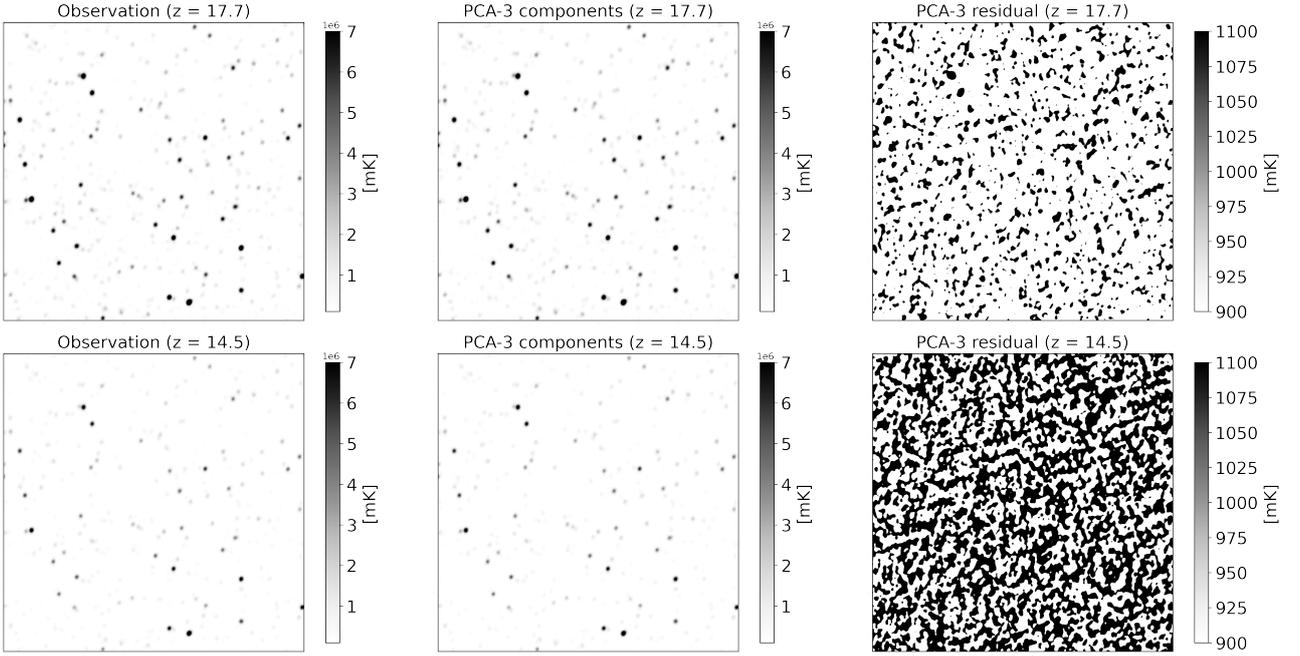}
 \caption{MWA G0044 project observation temperature map (left), PCA-3 components (middle) and the residual map after removing PCA-3 components (right) at redshift z = 17.7 (top) and z = 14.5 (bottom) at center frequency of 88 MHz. PCA-3 results are derived from the analysis of four sub-band images.}
\label{fig:pca3_87}
\end{center}
\end{figure*}

\begin{figure*} 
\begin{center}
 \includegraphics[width=17.0cm]{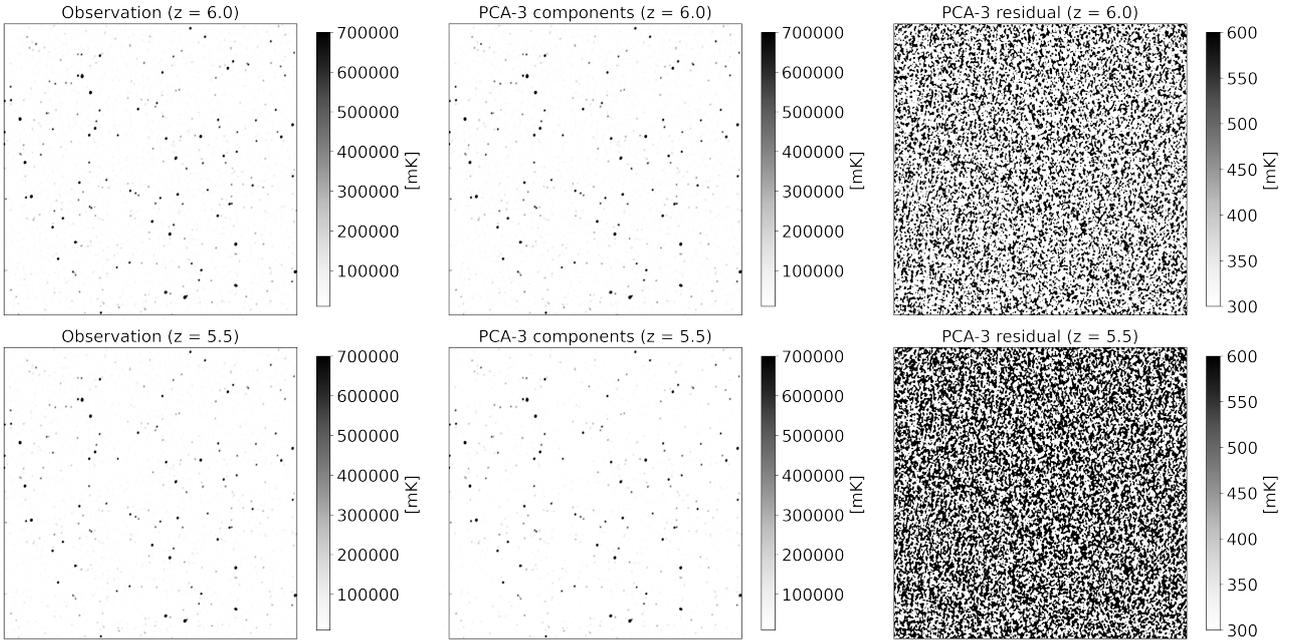}
 \caption{MWA G0044 project observation temperature map (left), PCA-3 components (middle) and the residual map after removing PCA-3 components (right) at redshift z = 6.0 (top) and z = 5.5 (bottom) at center frequency of 216 MHz. PCA-3 results are derived from the analysis of four sub-band images.}
\label{fig:pca3_215}
\end{center}
\end{figure*}

In our analysis, we use PCA provided by scikit-learn \citep{pedregosa2011scikit} as our foreground removal algorithm, and remove the first three principal components from our observation temperature map. Henceforth, the notation PCA-$N_{\rm comp}$ corresponds to signal for which the first $N_{\rm comp}$ principal components have been removed. The results at different sub-bands are shown in Figure \ref{fig:pca3_87} at 88 MHz and Figure \ref{fig:pca3_215} at 216 MHz. After analysing each PCA component, we believe that the first component represents the point sources. The number of radio sources in observation, PCA components, the matched sources between observation and PCA, and the ratio of the number of matched sources to observation detected in each sub-bands of two frequency-band are listed in Table \ref{tab:fg_number}. Thus we find that nearly 100\% of the radio sources in observation have been extracted by PCA-3. As shown in Figures \ref{fig:pca3_87} and \ref{fig:pca3_215}, that PCA-3 can extract the most foreground components such as point sources in the observation while the residual map includes noise and cosmological signal. 

\begin{table}
\centering
\caption{Comparison of number of radio sources detected before and after the Principal Component Analysis (PCA-3) foreground removal algorithm.}
\label{tab:fg_number}
\begin{tabular}{c|cccc}
\hline
z     & Observation & PCA-3 & Matched & Matched / Observation \\
\hline
17.65 & 319 & 319   & 319   & 100\%   \\
15.94 & 310 & 310   & 310   & 100\%   \\
14.52 & 328 & 378   & 324   & 98.78\% \\
13.32 & 327 & 346   & 322   & 98.47\% \\
\hline
5.96  & 551 & 553   & 549   & 99.64\% \\
5.71  & 547 & 547   & 546   & 99.82\% \\
5.47  & 525 & 613   & 523   & 99.62\% \\
5.25  & 504 & 641   & 503   & 99.80\% \\
\hline
\end{tabular}
\end{table}

\subsection{Power Spectrum Measurement}

The most significant cosmological parameter constraints from HI intensity mapping will likely arise from the power spectra of the 21cm brightness temperature, as two-point correlation functions encompass the majority of information pertaining to the underlying cosmology on large, linear scales. For this study, we focus on analysing angular power spectra, which captures clustering patterns on the sky.

For a specific frequency and assuming a full-sky survey, the angular power spectrum of the brightness fluctuations $\Delta T_{\rm b}$ is determined by computing the spherical harmonic components:
\begin{equation}
    a_{\ell m}(\nu) = \int \rm 
    d {\hat{\textbf{n}}}^2 \Delta T_{\rm b}(\nu, \hat{\textbf{n}}) Y_{\ell m}^*(\hat{\textbf{n}})
\end{equation}
where $Y_{\ell m}(\hat{\textbf{n}})$ are the spherical harmonic basis functions. We can then estimate the power spectrum by taking the average of the moduli of the harmonics components:
\begin{equation}
    \Tilde{C_{\ell}} = \frac{1}{2\ell + 1} \sum_{m = - \ell}^{\ell} |a_{\ell m}|^2
\end{equation}
where small values of $\ell$ correspond to the largest scales. Based on the healpy Python library \citep{zonca2019healpy}, we calculate the angular power spectra for our maps. The variance of the angular power spectra $\Delta C_\ell$ is approximated by:
\begin{equation}
    \Delta C_{\ell} = \sqrt{\frac{2}{(2 \ell + 1)f_{\rm sky}}} C_{\ell} 
\end{equation}
where $f_{\rm sky}$ = FOV / 129600 deg$^2$, and FOV = 4.9$^2$ deg$^2$ represents the area of our selected sky coverage.

\begin{figure*}
\begin{center}
 \includegraphics[width=17.0cm]{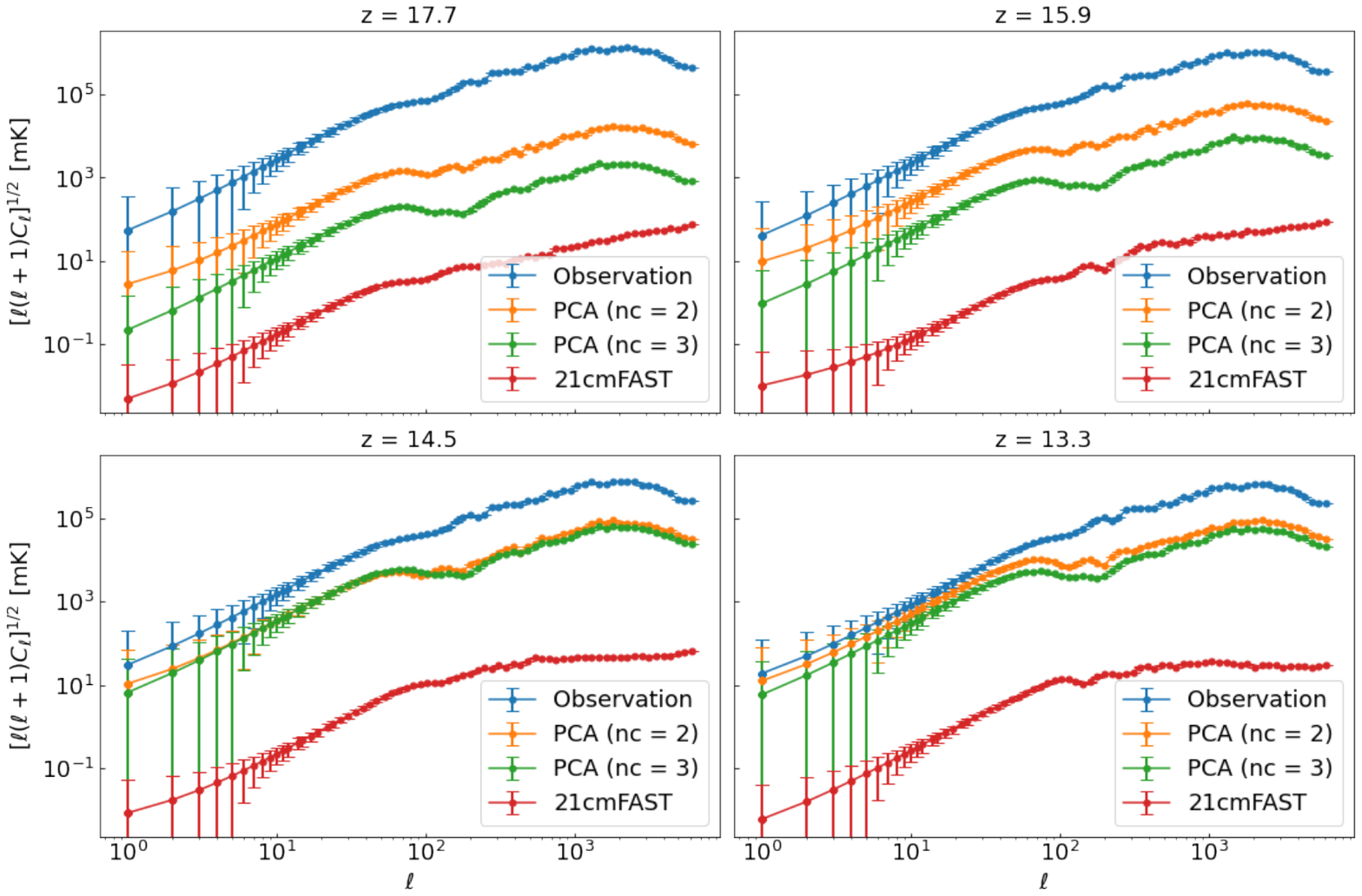}
 \caption{Angular power spectrum $C_\ell(0)$ at redshift 17.7 (top left), 15.9 (top right), 14.5 (bottom left), 13.3 (bottom right) for observation temperature map (blue), the residual map after removing 2 (orange) and 3 (green) PCA components and cosmological signal generated by 21cmFAST for comparison. PCA results are derived from the analysis of four sub-band images.}
\label{fig:cell_87}
\end{center}
\end{figure*}

After removing the foregrounds from each sub-band image, we utilize these four 7.68-MHz sub-band images, at two different frequency centers: 88 MHz and 216 MHz, for the measurement of the statistical angular power spectrum $C_\ell(\Delta\nu = 0)$. 
Figure \ref{fig:cell_87} shows the angular power spectrum at 88 MHz using the integration time of 338 minutes. Since we adopt the MWA Phase II extended configuration to collect the data, we force more on getting the foreground components, and the angular power spectrum of PCA components are very close to the input observation. After removing 2 or 3 PCA foreground components, the angular power spectrum of the residual map decreases in all $\ell$-modes in comparison to the input observation temperature map. However, it is still more than one magnitude higher than the theoretical estimated cosmological signal. Regarding the simulation of the EoR signal, we utilize the 2016 data release from the Evolution of Structure (EOS) project\footnote{\href{http://homepage.sns.it/mesinger/EOS.html}{http://homepage.sns.it/mesinger/EOS.html}} \citep{mesinger2016evolution}, which uses a publicly available code 21CMFAST \citep{mesinger2007efficient, mesinger201121cmfast}. From the light-cone cube, we extract image slices corresponding to specific sub-band frequencies. These extracted image slices are subsequently re-scaled to align with the sky coverage (FOV = 24 degrees$^2$) and pixel size of the deep images at two frequency-band. 

\begin{figure*}  
\begin{center}
 \includegraphics[width=17.0cm]{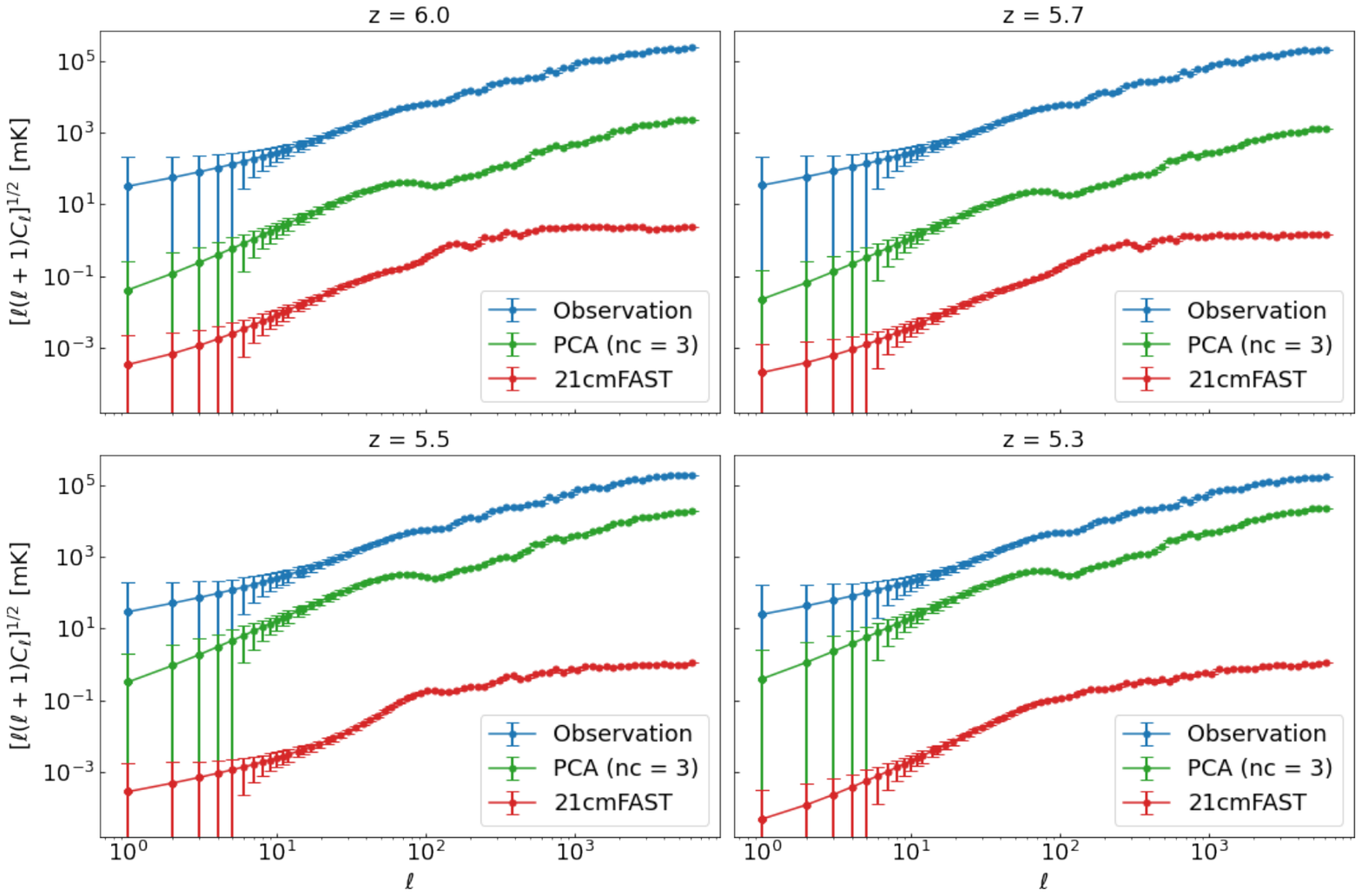}
 \caption{Angular power spectrum $C_\ell(0)$ at redshift 6.0 (top left), 5.7 (top right), 5.5 (bottom left), 5.3 (bottom right) for observation temperature map (blue), the residual map after removing 3 (green) PCA components and cosmological signal generated by 21cmFAST for comparison. PCA results are derived from the analysis of four sub-band images.}
\label{fig:cell_215}
\end{center}
\end{figure*}

Figure \ref{fig:cell_215} shows the angular power spectrum at 216 MHz using the integration time of 266 minutes. Similarly, after removing three PCA foreground components, the angular power spectrum of the residual map decreases in all $\ell$-modes in comparison to the input observation temperature map. However, it is still more than one magnitude higher than the theoretical estimated cosmological signal generated by 21cmFAST.

On the other hand, we notice that a local minima appears in both PCA-2 and PCA-3 in all sub-bands at 88 MHz. Specifically, for PCA-3, the local minima occurs at $\ell$ = 178 in the angular power spectrum of all sub-bands of its residual map. In contrast, for PCA-2, the local minima in the angular power spectrum of its corresponding four sub-bands are found at $\ell$ = 105, 104, 94 and 115. Additionally, at 216 MHz, PCA-3 exhibits a local minimum at $\ell$ = 116 in all its sub-bands. This might be due to incomplete removal of power from diffuse foregrounds, which cover the entire visible radio sky and may introduce duplicate flux contributions at higher $k$-modes (smaller angular scales) \citep{line2025verifying}.

\section{Conclusions and Discussion}
\label{sec:discussion}

Deep imaging of the structures of the Epoch of Reionization (EoR) across five targeted fields has been identified as one of the highest-priority scientific objectives for SKA1-Low. Selecting candidate fields with low average surface brightness, minimal variance, and other favourable characteristics is a crucial step in preparing future high-quality deep imaging. Building on the work of \citet{zheng2020pre}, we utilise data observed with the MWA Phase II extended configuration to investigate the properties of a selected ‘quiet’ field. A comprehensive understanding of the properties of radio sources, such as their flux distribution and source classification, is essential for developing accurate sky models for the data reduction pipeline and for algorithms aimed at foreground subtraction.

We provide a data reduction pipeline that encompasses the entire process, from data conversion to calibration and deep imaging. To investigate the statistical properties of radio sources, we select a circular region with a 5-degree radius, centred at [(RA, Dec) = (8:00:00, +5:00:00)], and employ AEGEAN to identify radio sources. Data from two frequency bands, 72–103 MHz and 200–231 MHz, are considered, each with a bandwidth of 30.72 MHz and centred at 88 MHz and 216 MHz, respectively. Four different integration times are analysed: 10, 20, 80, and 338 minutes at 88 MHz, and 10, 20, 80, and 266 minutes at 216 MHz, which are used to evaluate source counts and noise performance.

We have deduced various source counts (Integrated source counts, Differential source counts and Euclidean-normalized differential source counts) within our selected circular region with different integration times at two frequency-band. Specifically, at 216 MHz with integration time of 266 minutes, a maximum number of 2,576 radio sources were detected in G0044, and in the same sky region a total number of 971 sources detected in GLEAM catalogue are all matched with source detected in G0044. Generally, more radio sources are found by AEGEAN with increasing integration time at two frequency-band. Based on the single power law fits, we calculate the best-fit parameters $k=2866 \pm 442, \gamma=1.96 \pm 0.10$ at 88 MHz with integration time of 338 minutes for flux greater than 49 mJy, while $k= 2907 \pm 219, \gamma=1.67 \pm 0.03$ at 216 MHz with integration time of 266 minutes for flux greater than 10 mJy. We then use data with the maximum integration time to generate a matched source catalogue based on data from 216 MHz, which has a better angular resolution than the image at 88 MHz. 

Among the detected sources, 15.0\% and 16.5\% are classified as extended sources at 88 MHz and 216 MHz. No sources are excluded in the false detection rate (FDR) process. We estimate that our sources is 90\% complete above 81 mJy at 88 MHz and 10.4 mJy at 216 MHz, as are listed in Table \ref{tab:catalogue_properties}. Based on the results of FDR and source completeness, we derive the final corrected normalised source counts ($S^{2.5} $d$N/$d$S$) presented in Table \ref{tab:corrected_endc}.

The noise analysis results, including RMS, thermal noise, and confusion noise, are presented in Table \ref{tab:noise_estimates}. Confusion noise is calculated based on the source counts derived from deep images of the selected field observed with the MWA Phase II extended configuration. According to our analysis, a deep image with an integration time of 266 minutes at 216 MHz achieves a confusion noise level of 0.17 mJy. In conclusion, for future deep imaging of the Epoch of Reionisation (EoR) with SKA1-Low, a sky field similar to the G0044 field could achieve a lower confusion noise than that observed with the MWA. Future SKA1-Low will have better sensitivity and resolution than MWA. By employing long integration times, SKA1-Low observations can reduce confusion noise to levels below the thermal noise threshold.

We employ a foreground removal algorithm, PCA-3, and find that 98\% of the radio sources in the selected square region (FOV = 4.9 degrees) are successfully extracted by PCA-3. After removing the PCA-3 foreground components, the angular power spectrum of the residual map shows a decrease across all $\ell$-modes  compared to the input observation temperature map. However, it remains more than an order of magnitude higher than the theoretical cosmological signal predicted by 21cmFAST.

For the next steps, we plan to expand our analysis to cover a broader sky area and a wider range of frequencies, in addition to the foreground removal analysis. We have already conducted observations using the MWA across two fields and five frequency bands. In this paper, we focus on processing data from one field at two frequency bands, but in the future, by analysing these selected sky fields, we aim to deepen our understanding of foreground properties, refine our sky models, and test our analysis pipelines. Furthermore, we hope to extend our work to extracting large-scale structures from data obtained using the MWA's compact configuration, with the long-term goal of processing data from SKA1-Low with improved accuracy.

To achieve these objectives, improving the data reduction and foreground subtraction techniques is crucial. In particular, advancements in beam model characterisation and the application of more sophisticated algorithms will play an important role. \citet{nasirudin2022characterizing} uses the OSKAR simulator \citep{mort2010oskar} to simulate the SKA beam, leveraging PCA and Kernel PCA techniques to characterise the beam uncertainties. This characterisation is crucial for reducing errors in the power spectrum, which are essential for precision measurements in radio astronomy. Additionally, \citet{li2019separating} simulates observation data based on the SKA1-Low layout configuration using the same OSKAR simulator. In their study, they employ a convolutional denoising autoencoder (CDAE), a deep-learning-based foreground removal algorithm, to address the complex beam effects and accurately separate the EoR signal. \citet{ni2022eliminating} simulated two beam models: the Gaussian beam model and the Cosine beam model. The PCA+U-Net foreground subtraction method can effectively eliminate the telescope primary beam effect, providing new insights into recovering the HI power spectrum for future observations. Therefore, in future work related to preparing observations of these deep fields with SKA1-Low, we will simulate the SKA1-Low beam and assess the impact of the beam structure on foreground subtraction algorithms such as PCA. This will help us determine whether these algorithms can extract EoR signals more accurately, thus contributing to the data reduction pipeline for SKA1-Low EoR observations and improving accuracy.

Besides of better characterising the beam model, further advances can be made by improving data reduction methods and foreground subtraction techniques. \texttt{hyperdrive}\footnote{\href{https://github.com/MWATelescope/mwa_hyperdrive}{https://github.com/MWATelescope/mwa\_hyperdrive}} is a calibration software developed for the MWA radio telescope. Fast Holographic Deconvolution (FHD) is an open-source imaging algorithm designed for radio interferometers \citep{sullivan2012fast, barry2019fhd}, which has been tested on the MWA and is primarily focused on compact sources. Additionally, Python Fast Holographic Deconvolution (pyFHD)\footnote{\href{https://github.com/ADACS-Australia/PyFHD}{https://github.com/ADACS-Australia/PyFHD}}, an improved Python version of FHD, is currently under development. We also plan to incorporate ionospheric corrections into our calibration pipeline to improve the quality of our data reduction output. To further enhance the results from our foreground removal algorithm, we plan to explore improved methods based on PCA or deep learning approaches in the future.

\section*{Acknowledgements}

This work is supported by National SKA Program of China No.SQ2020SKA0110100 and No.SQ2020SKA0110200.
We thank Cathryn Trott for reading the manuscript and providing many helpful comments for this paper. We thank the anonymous referee for their insightful suggestions that have improved this paper. This scientific work uses data obtained from Inyarrimanha Ilgari Bundara / the Murchison Radio-astronomy Observatory. We acknowledge the Wajarri Yamatji People as the traditional owners and native title holders of the Observatory site. Establishment of CSIRO's Murchison Radio-astronomy Observatory is an initiative of the Australian Government, with support from the Government of Western Australia and the Science and Industry Endowment Fund. Support for the operation of the MWA is provided by the Australian Government (NCRIS), under a contract to Curtin University administered by Astronomy Australia Limited. This work was supported by resources provided by the Pawsey Supercomputing Research Center with funding from the Australian Government and the Government of Western Australia.

\addcontentsline{toc}{section}{Acknowledgements}

\section*{Data Availability Statement}
The data used in this work are available from the MWA archive via https://asvo.mwatelescope.org/.



\bibliographystyle{mn2e}
\bibliography{ref}




\appendix
\section{Catalogue}    

\begin{landscape}
\begin{table}
\centering
\caption{Source catalogue within 5-degree radius (top 20 sorted by flux density at 216 MHz).}
\label{tab:source_cat}
\begin{tabular}{ccccccccccccccccc}
\hline
RA & Dec & flux\_88 & err\_flux\_88 & rms\_88 & a\_88 & b\_88 & pa\_88 & flux\_216 & err\_flux\_216 & rms\_216 & a\_216 & b\_216 & pa\_216 & classification & alpha & err\_alpha \\
{[}h:m:s{]} & {[}d:m:s{]} & {[}Jy{]} & {[}Jy{]} & {[}Jy/beam{]} & {[}arcsec{]} & {[}arcsec{]} & {[}degree{]} & {[}Jy{]} & {[}Jy{]} & {[}Jy/beam{]} & {[}arcsec{]} & {[}arcsec{]} & {[}degree{]}\\
\hline
07:45:04.54 & +02:00:04.79 & 18.8078 & 0.0017 & 0.0075 & 288.5952 & 167.1365 & -17.3042 & 10.1105 & 0.0010 & 0.0026 & 210.2416 & 77.1069  & -15.4578 & point-like & -0.6908 & 0.0006 \\
07:55:45.86 & +02:10:38.15 & 12.3017 & 1.2302 & 0.0088 & 249.9081 & 169.6920 & -16.9263 & 5.8368  & 0.0005 & 0.0018 & 126.4872 & 83.4359  & -14.2326 & point-like & -0.8802 & 0.0267 \\
08:15:22.87 & +01:55:03.66 & 18.0890 & 0.0027 & 0.0109 & 245.1486 & 166.5525 & -17.0125 & 5.6815  & 0.5682 & 0.0018 & 120.6227 & 75.2107  & -14.9541 & point-like & -1.3381 & 0.0978 \\
07:42:39.32 & +05:07:04.28 & 5.6607  & 0.0037 & 0.0089 & 257.2832 & 167.5008 & -17.2862 & 3.1345  & 0.3134 & 0.0024 & 125.6434 & 79.8289  & -15.1286 & point-like & -0.8888 & 0.0153 \\
07:57:41.53 & +02:51:00.74 & 4.7005  & 0.0080 & 0.0092 & 257.4870 & 173.6102 & -17.6614 & 2.3610  & 0.0003 & 0.0017 & 121.1359 & 74.8686  & -14.9993 & point-like & -0.7653 & 0.0026 \\
08:00:01.43 & +02:22:41.70 & 4.2394  & 0.4239 & 0.0080 & 245.0276 & 164.6352 & -17.1540 & 2.0854  & 0.0004 & 0.0015 & 120.9203 & 75.6559  & -15.0620 & point-like & -0.7217 & 0.0095 \\
08:13:23.69 & +07:34:05.34 & 3.7825  & 0.0024 & 0.0076 & 259.5520 & 166.8706 & -16.8677 & 2.0837  & 0.0004 & 0.0021 & 126.3393 & 75.6098  & -14.7761 & point-like & -0.6637 & 0.0012 \\
08:04:34.39 & +06:33:43.90 & 3.0353  & 0.0011 & 0.0061 & 263.4253 & 167.0486 & -16.8528 & 2.0559  & 0.0006 & 0.0018 & 131.4819 & 76.5587  & -14.7403 & point-like & -0.4327 & 0.0016 \\
07:54:21.80 & +07:38:52.81 & 3.4536  & 0.0011 & 0.0074 & 267.4343 & 166.1093 & -17.0990 & 1.9367  & 0.0008 & 0.0018 & 127.7118 & 75.2013  & -14.8598 & extended   & -0.6438 & 0.0012 \\
08:00:43.81 & +08:31:32.65 & 3.1853  & 0.0012 & 0.0080 & 266.7084 & 165.7835 & -17.0057 & 1.7632  & 0.1763 & 0.0020 & 129.4752 & 75.1910  & -14.8525 & point-like & -0.7776 & 0.0102 \\
08:04:00.69 & +06:17:21.71 & 2.4019  & 0.0044 & 0.0060 & 258.8728 & 165.4002 & -16.9571 & 1.6764  & 0.0001 & 0.0018 & 125.4137 & 74.8829  & -14.9019 & point-like & -0.3983 & 0.0069 \\
07:58:37.50 & +08:01:48.17 & 2.5792  & 0.0161 & 0.0078 & 266.7590 & 174.3891 & -16.8426 & 1.5700  & 0.0002 & 0.0019 & 128.1183 & 85.6178  & -14.5892 & point-like & -0.5401 & 0.0176 \\
07:50:18.52 & +03:16:41.68 & 2.7671  & 0.0051 & 0.0071 & 251.0468 & 166.6114 & -17.2396 & 1.5223  & 0.1522 & 0.0018 & 123.3017 & 79.0768  & -15.3393 & extended   & -0.8300 & 0.0163 \\
07:58:41.75 & +03:54:41.40 & 2.5059  & 0.0017 & 0.0088 & 251.3929 & 164.4352 & -17.0475 & 1.4972  & 0.0008 & 0.0016 & 122.3511 & 74.4541  & -15.0054 & point-like & -0.5729 & 0.0019 \\
08:08:43.91 & +05:50:36.23 & 2.1210  & 0.0035 & 0.0068 & 257.8748 & 171.3840 & -16.7866 & 1.4899  & 0.1490 & 0.0018 & 129.2368 & 85.3080  & -14.3965 & point-like & -0.4738 & 0.0202 \\
08:07:57.64 & +04:32:33.70 & 3.3142  & 0.3314 & 0.0077 & 251.1412 & 164.8743 & -16.9845 & 1.4703  & 0.0014 & 0.0017 & 122.7134 & 74.8100  & -14.9233 & extended   & -0.8461 & 0.0179 \\
07:54:17.15 & +03:44:33.33 & 3.1138  & 0.0018 & 0.0075 & 252.8510 & 165.5216 & -17.1879 & 1.4689  & 0.1469 & 0.0017 & 123.5138 & 75.6140  & -15.0298 & point-like & -0.9453 & 0.0147 \\
08:11:05.60 & +04:20:18.05 & 0.9463  & 0.0008 & 0.0075 & 253.9808 & 169.4185 & -18.0032 & 1.4556  & 0.1456 & 0.0018 & 122.1329 & 74.7400  & -14.9022 & point-like & inf     & inf    \\
08:17:15.51 & +07:08:49.52 & 2.9604  & 0.2960 & 0.0083 & 296.7540 & 205.7897 & -15.6790 & 1.3611  & 0.0010 & 0.0021 & 138.9872 & 137.8076 & 77.4450  & extended   & -0.6804 & 0.0302 \\
07:56:07.71 & +01:07:37.77 & 1.9880  & 0.0114 & 0.0071 & 244.9964 & 181.7675 & -17.3260 & 1.2789  & 0.0001 & 0.0016 & 122.8034 & 112.4937 & -17.4043 & point-like & -0.4871 & 0.0114 \\
\hline
\end{tabular}
\end{table}

\begin{table}
\caption{Column numbers, names, and units for the catalogue.}
\label{tab:column_name}
\begin{tabular}{llcl}
\hline
Number & Name              & Unit       & Description               \\
\hline
1      & RA                & h:m:s      & Right Ascension (J2000)   \\
2      & Dec               & d:m:s      & Declination (J2000)       \\
3      & flux\_88          & Jy         & Integrated flux density of the source in 88 MHz image                         \\
4      & err\_flux\_88     & Jy         & Uncertainty in fit for integrated flux density of the source in 88 MHz image  \\
5      & rms\_88           & Jy/beam    & Local noise level in 88 MHz image                                             \\
6      & a\_88             & arcsec     & Fitted semi-major axis of the source in 88 MHz image                          \\
7      & b\_88             & arcsec     & Fitted semi-minor axis of the source in 88 MHz image                          \\
8      & pa\_88            & degree     & Fitted position angle of the source in 88 MHz image                           \\
9      & flux\_216         & Jy         & Integrated flux density of the source in 216 MHz image                        \\
10     & err\_flux\_216    & Jy         & Uncertainty in fit for integrated flux density of the source in 216 MHz image \\
11     & rms\_216          & Jy/beam    & Local noise level in 216 MHz image                                            \\
12     & a\_216            & arcsec     & Fitted semi-major axis of the source in 216 MHz image                         \\
13     & b\_216            & arcsec     & Fitted semi-minor axis of the source in 216 MHz image                         \\
14     & pa\_216           & degree     & Fitted position angle of the source in 216 MHz image                          \\
15     & classification    & -          & Source classification in 216 MHz image                                        \\
16     & alpha             & -          & Fitted spectral index                                                         \\
17     & err\_alpha        & -          & Error on fitted spectral index \\
\hline
\end{tabular}
\end{table}

\end{landscape}


\bsp	
\label{lastpage}
\end{document}